\documentclass[fleqn,usenatbib]{mnras}

\usepackage{newtxtext,newtxmath}

\usepackage[T1]{fontenc}

\DeclareRobustCommand{\VAN}[3]{#2}
\let\VANthebibliography\thebibliography
\def\thebibliography{\DeclareRobustCommand{\VAN}[3]{##3}\VANthebibliography}

\usepackage{graphicx}	
\usepackage{amsmath}	

\usepackage[table]{xcolor}

\title[Dating stellar populations with extended HB]{Deriving ages and horizontal branch properties of integrated stellar populations}

\author[I. Cabrera-Ziri]{
Ivan Cabrera-Ziri,$^{1,2,3}$\thanks{E-mail: cabrera@uni-heidelberg.de (ICZ)} 
Charlie Conroy$^{3}$
\\
$^{1}$ Astronomisches Rechen-Institut, Zentrum f\"ur Astronomie der Universit\"at Heidelberg, M\"onchhofstra{\ss}e 12-14, D-69120 Heidelberg, Germany\\
$^{2}$Astrophysics Research Institute, Liverpool John Moores University, 146 Brownlow Hill, Liverpool L3 5RF, UK\\
$^{3}$Center for Astrophysics | Harvard \& Smithsonian, 60 Garden Street, Cambridge, MA 02138, USA\\
}

\date{Accepted 2021 December 31. Received 2021 December 7; in original form 2021 May 18}

\pubyear{2021}

\begin{document}
\label{firstpage}
\pagerange{\pageref{firstpage}--\pageref{lastpage}}
\maketitle

\begin{abstract}
A major source of uncertainty in the age determination of old ($\sim10$ Gyr) integrated stellar populations is the presence of hot horizontal branch (HB) stars.
Here, we describe a simple approach to tackle this problem, and show the performance of this technique that simultaneously models the age, abundances and HB properties of integrated stellar populations. For this we compare the results found during the fits of the integrated spectra of a sample of stellar population benchmarks, against the values obtained from the analysis of their resolved CMDs. We find that the ages derived from our spectral fits for most (26/32) of our targets are within $0.1 {\rm ~dex}$ to their CMDs values. Similarly, for the majority of the targets in our sample we are able to recover successfully the flux contribution from hot HB stars (within $\sim0.15 {\rm ~dex}$ for 18/24 targets) and their mean temperature (14/24 targets within {$\sim30 \%$}). Finally, we present a diagnostic that can be used to detect spurious solutions in age, that will help identify the few cases when this method fails. These results open a new window for the detailed study of globular clusters beyond the Local Group.
\end{abstract}




\section{Introduction}
\label{sec:intro}

Obtaining accurate ages of extragalactic globular clusters would provide a much needed constraint on several astronomical problems including reionization, galaxy assembly as well as globular cluster formation itself \citep[see][]{Forbes18}. The reason for this is that globular clusters are expected to form at different epochs depending of the mechanism at play in their formation.  For example, scenarios where globular clusters form in mini dark matter haloes \citep[e.g.][]{Griffen10,Trenti15}, preferentially produce globular clusters during/before reionization ($\sim13 \mbox{~Gyr}$ ago at $z\sim7$). On the other hand, scenarios where globular clusters are the product of star formation in high pressure/density environments \citep[e.g.][]{Muratov10,Kruijssen15}, expect that most globular clusters formed closer to the bulk of the stars in galaxies ($\sim10.5\mbox{~Gyr}$ ago at $z\sim2$, e.g. \citealt{Madau14}). Both scenarios have interesting implications.

If globular clusters formed very early in the Universe they could have had an important contribution during reionization  \citep[e.g.][]{He20,Ma21}. Similarly, if observations were to suport the formation of globular clusters in mini-haloes, they could be used to understand the nature of dark matter in small scales. On the other hand, if globular clusters were to be the outcome of (vigorous) star formation, these systems could help us track the star formation process and assembly history of galaxies \citep[e.g.][]{Reina-Campos17,Kruijssen19}.

The precision in relative age obtained from the analysis of the colour-magnitude diagrams (CMDs) of Galactic globular clusters, in combination with state-of-the-art models, has started being used for some of the applications discussed above \citep[e.g.][]{Kruijssen19}. However, large samples of globular clusters from different galaxies are required to reach robust conclusions, as significant variance in the mean age (and dispersion) of the globular clusters systems is expected for different galaxies according to some models \citep[e.g.][]{Kruijssen19a}. The problem is that CMD ages can be obtained only for selected nearby targets like the Magellanic Clouds and other dwarf satellites \citep[e.g.][]{Glatt08,Glatt09,deBoer16}, and although they are not impossible these observations already become extremely expensive at the distances of M 31 \citep[e.g.][who required 120 \emph{HST} orbits to reach the main sequence turn-off of \emph{one} globular cluster]{Brown04}. Although facilities like \emph{JWST} are expected to improve the efficiency of these observations, for targets beyond the Local Group the ages will have to be inferred from the properties of their integrated light.

Unfortunately, it has been long known that the ages of old ($\sim10$ Gyr) stellar populations derived from the analysis of integrated light can be significantly biased towards younger values by the presence of hot horizontal branch (HB) stars 
\citep[e.g.][]{Worthey94,Freitas95,Beasley02,Thomas05}. This represents a major source of systematic error in the determination of ages \citep[the largest one for sub-solar metallicities, e.g.][]{Ocvirk10}. To understand the origin of this problem one needs to understand what drives the properties of HB stars and how these parameters are included in standard stellar evolution models (i.e. isochrones) used in the synthesis of stellar populations.

Different parameters influence the characteristics of HB stars like the mass of the star (and He core) at the time it arrives to the HB and chemical composition \citep{Cassisi13}. Even stellar populations of the same age and metallicity can manifest different HBs, most likely reflecting differences in mass loss during the red giant branch phase (both the mean value and its scatter) and/or the initial He abundance (spread).

The issue is that some of these parameters cannot be predicted from current stellar evolution theory. Standard stellar population models simply adopt a set of values for these parameters, e.g. mass loss efficiency $\eta$ (with no scatter), He abundance that is scaled to the metallicity (again, without a scatter), etc\footnote{Although see e.g. \cite{Lee00,Maraston05,Percival11,Chung17} for stellar populations models that allow more flexibility. }. Although these simplifications are justified, by fixing these values one necessarily compromises the ability to describe the properties of stellar populations with extended HBs, i.e. where HB stars extend to hotter temperatures than the predicted from standard models\footnote{In the literature these are often referred to as blue HB or BHB, but here we avoid this term as it can be confused with `Blue Hook' stars (which are HB with extremely hot temperatures). The reason for the distinction is that not all extended HB necessarily reach the extreme temperatures of Blue Hook stars.}.

The lack of flexibility in these parameters has as a consequence that when we use standard models to analyse an old stellar population hosting an extended HB, we will often require a young age to reproduce their integrated light -- both SED (colours) and spectrum (e.g. Balmer lines). The reason is that the only hot stars that can mimic the luminosity/temperatures of the extended HB stars included in such models are stars of younger age (main sequence stars), hence the bias in age.

Over the years there have been different approaches to deal with this issue. For example, \cite{Schiavon04} introduced a diagnostic based on the ratio of Balmer lines that was capable of separating objects with different HB morphologies. Alternatively, \cite{Percival11} proposed the use of \ion{Mg}{ii} and \ion{Ca}{ii} features to identify the presence of hot HB stars in the spectra of old stellar populations. Another step forward was presented by \cite{Koleva08}, who added to their standard stellar population models, the contribution of a linear combination of hot stars (between $6-20 {\rm ~kK}$). \citeauthor{Koleva08} found a significant improvement in the age recovery of old stellar populations with extended HB stars with their approach \citep[see figures 6 and 7 in][]{Koleva08}.
However, despite all this, the presence of extended HB stars in old stellar populations remains a severe limitation in the age determination of unresolved stellar populations, even when using state-of-the-art tools e.g. \cite{Perina11,Conroy18,Goncalves20,Johnson20}. 

In this work, we test the implementation of extended HB stars in stellar populations models used in the analysis of integrated spectra with an approach similar in spirit to \cite{Koleva08}, with the goal to use it in future analyses of extragalactic globular clusters. For our experiments we use Galactic globular clusters as benchmarks, since there is available reliable determinations of their ages, HB properties and [Fe/H] from independent techniques -- namely CMDs and high-resolution spectra of individual stars.

This work is structured as follows, in Section \ref{sec:alf} we describe our modelling technique. The details of the data used for our experiments can be found in Section \ref{sec:data} and some relevant properties of our sample in \ref{sec:ref}. Our main results are presented in Section \ref{sec:res}, these are discussed in more detail in Section \ref{sec:disc} and in Section \ref{sec:fut} we discuss some areas that could be improve in future studies. Finally in Section \ref{sec:sum} we provide a summary with conclusions.

\section{Modelling the integrated spectra}
\label{sec:alf}

The fits to the integrated spectra were carried out with \texttt{alf} \citep[see][]{Conroy12,Conroy14,Choi14,Conroy18}, which is capable of modelling the optical to near IR spectrum (0.37--2.4 \micron) of old ($>1 {\rm Gyr}$) stellar populations over a wide metallicity range, approximately from -2.0 to +0.3. The models used in the spectral fits are very versatile and include parameters like: the recession velocity and velocity dispersion, age, overall metallicity, abundances of some individual elements, the slopes of the IMF as well as the age and mass fraction of a younger sub-population. Discussions about implementation and the performance of the recovery of these parameters can be found in the references listed above. In the rest of this section we focus on describing some updates regarding the modelling of the extended HB in \texttt{alf}.

As mentioned in Section \ref{sec:intro}, the simplifications used in construction of standard isochrones can introduce limitations when modelling the HB of some stellar populations. In our context, this means a shortfall when describing the hot end of the $T_{\rm eff}$ distribution of HB stars. In order to compensate for this effect, \texttt{alf} has the option to include the flux contribution of a hot HB star \emph{over and above} the prediction from a standard isochrone. The model for this additional hot HB star has two parameters: the temperature of the hot star, $T_{\rm hot}$, and its contribution to the flux of the stellar population, $f_{\rm hot}$. Specifically, the later is implemented as the ($\log$) fraction of flux of the hot star to the entire spectrum at 0.5\micron~ for this analysis.

During the fits the spectrum of the hot star component is obtained from the interpolation of hot HB spectra templates with ${\rm[Fe/H]}={-1.50,-1.00,-0.50,0.00,0.25} {\rm ~dex}$ and temperatures $8 \leq T_{\rm eff} {\rm ~(kK)} \leq 30$ in intervals of $2 {\rm ~kK}$. These templates were calculated using the same set of assumptions as the standard theoretical spectral library used in \texttt{alf}; see \citet{Conroy18} for details.

For this work, we will focus on the region between 3750--5500 \AA~where the presence/absence of an extended HB population has a significant influence of the integrated spectrum of stellar populations. In our fits we assumed uniform priors in $\log({\rm age / Gyr}) = (\log(0.5),\log(14))$, ${\rm[M/H]}=(-2,0.3)$, $\log(f_{\rm hot})=(-6,0)$ and $T_{\rm hot} {\rm(kK)}=(8,30)$ and a fixed \cite{Kroupa01} initial mass function. 

The parameter space was sampled with \texttt{emcee} \citep{Foreman-Mackey13} with 1024 walkers. The burn-in stage was modified to reset the walkers after $10^5$ steps in a position near the maximum likelihood in order to speed up the convergence. Then the burn-in resumes and after another $10^5$ steps a production run begins for the final posterior distribution.

For convenience/simplicity, from now on we will refer as `standard models' the stellar populations built from the MIST isochrones  version 1.2   \citep[for more details, see][]{Choi16}. These standard models adopt a constant initial He composition at a given metallicity and a RGB mass loss following \cite{Reimers75} with $\eta = 0.1$ (with no scatter), which cannot reproduce the entire variety of HB morphologies observed in globular clusters.

\section{Data}
\label{sec:data}

For our analysis of Galactic globular clusters we use the integrated spectra presented in \cite{Schiavon05}. This library contains the spectra of 41 globular clusters covering $\sim 3300-6500$ \AA~at high SNR ($> 100 \mbox{~\AA}^{-1}$ at 5000 \AA). For some targets, different exposures were taken and in some cases the spectra was extracted using different apertures. Here, we analyse the spectra taken during the exposure `a' and extracted using aperture `1' for all our targets.

Since the spectra of \citeauthor{Schiavon05} has a higher spectral resolution than our models $(100\,{\rm km\,s^{-1}})$, we downgraded its resolution to $150\,{\rm km\,s^{-1}}$ before we analysed the spectra. We also masked a few regions where the data seemed to be corrupted, namely 4145--4170 \AA, 4303--4322 \AA, 4520--4568 \AA~and 5025--5068 \AA.

One of the goals of our analysis is to simulate the (mean) properties of the extended HB population of these clusters and its contribution to their integrated spectra. To evaluate the performance of our technique, we carried out independent measurement of these parameters using the colour magnitude diagram obtained from high resolution \emph{HST} photometry of the \citeauthor{Schiavon05} targets (see Section \ref{sec:phot_hb}). For this we used the publicly available catalogues from the WFPC2 \citep{Piotto02} and ACS \citep{Sarajedini07} Galactic globular cluster surveys.

Of the Galactic globular clusters from \citeauthor{Schiavon05}, 9 did not have good/clean \emph{HST} CMDs, so we exclude them from our analysis. Most of these targets are located in the Galactic Bulge, where the background/foreground density of stars is high resulting in severe CMD contamination.

Since the Galactic globular cluster population is predominantly old ($\gtrsim10$ Gyr), we decided to complement it with a sample of younger globular clusters from the LMC/SMC in order to study the performance of our technique in a wider range of ages (with clusters as young as $\sim 1$ Gyr). For this we analysed the spectra of 7 LMC/SMC clusters that had CMD ages reported in the literature $<9$ Gyr (see Table \ref{tab:lit}).

The integrated spectra for the younger sample came from the WAGGS survey \citep[][]{Usher17}. This survey targeted $>120$ globular clusters of the Milky Way and its satellites to obtain spectra at higher resolution (FWHM of 0.8 \AA~vs. 3.1 \AA) and wavelength coverage ($\sim3300-9050$ \AA) than \cite{Schiavon05}. However, the SNR of the spectra in the bluest regions is significantly lower than the ones from the \citeauthor{Schiavon05} dataset  \citep[most with SNR $<25$ for $\lambda<4350$ \AA ,~see][]{Goncalves20}, so we preferred the latter for our analysis of Galactic targets. Like for the \citeauthor{Schiavon05} data set we downgraded the resolution of the young globular cluster spectra to $150\,{\rm km\,s^{-1}}$ before our fits. We found no need to mask the WAGGS spectra.

In summary, our final sample is constituted by 39 clusters, where 7 are young clusters from the LMC/SMC and 32 are old Galactic clusters. Of the latter, 24/32 have an extended HB population.

\section{Reference vaules}
\label{sec:ref}
\subsection{Ages, metallicities, distances and reddening}
\label{sec:phot_ages}

For each cluster in our sample we will compare the inferred age, [Fe/H] and extended HB properties with the ones reported in the literature from the analysis of resolved photometry and spectroscopy of cluster stars.

For Galactic globular clusters the reference ages and [Fe/H] come from \cite{Forbes10,Dotter10,Dotter11,VandenBerg13}. Following \cite{Kruijssen19}, for clusters with values reported in more than one of these works, we took as the reference value the mean value between different works and its uncertainty the standard deviation. The distances used for most clusters are the ones reported in \cite{Baumgardt19} and the reddening is from \cite{Harris96} (2010 edition). That said, in a few cases we found a better agreement using alternative values, those instances are also reported in Table \ref{tab:lit}. 
In this table we also report the values for the young LMC/SMC clusters included in our analysis and their respective references.

\subsection{Extended HB parameters from CMD}
\label{sec:phot_hb}

The stellar populations used in \texttt{alf} are built using the MIST isochrones \citep{Choi16}. As discussed above, at the metallicities and ages of interest the MIST grid does not produce particularly hot HB stars (see Fig. \ref{fig:cmd} below). To account for this, in \texttt{alf} we model the extended HB population by adding to the standard stellar population model spectrum the contribution of a hot star representing the mean properties of the extended HB population. Here we describe how we measure the parameters of this extended (hot) component, namely the fraction of flux (in $\log$ units) to the integrated light coming from extended HB stars, $\log (f_{\rm hot})$, and their mean $T_{\rm eff}$, $T_{\rm hot}$, using the \emph{HST} CMD of the clusters.

First we identify the extended HB population in each of the cluster CMDs by visual inspection. Then we add up the flux of all this stars in $F555W$/$F606W$ for targets with WFPC2/ACS data respectively, which is then divided by the flux of \emph{all} the stars in the CMD in the same band. We take the logarithm of this value to obtain $\log (f_{\rm hot})$.

To calculate $T_{\rm hot}$, we take the light weighted mean $T_{\rm eff}$ of extended HB stars (in the $F555W$ or $F606W$ band depending of the cluster). These temperatures are inferred using the \cite{Pietrinferni06} colour-${\rm T}_{\rm eff}$ relation for zero age horizontal branch stars (with mass loss according to \citeauthor{Reimers75} law and $\eta = 0.4$), after accounting for the distance and reddening.

Finally, as mentioned in Section \ref{sec:alf}, the parametrization in \texttt{alf} is such that the different fractions of flux for the hot HB star during our MCMC sampling are drawn at 0.5\micron. So in order to provide a more appropriate comparison of the extended HB flux derived from our spectral fits to the ones inferred from the CMDs, we recalculate the contributions of the extended HB component inferred from our spectral fits in the $F555W$ and $F606W$ bands.

\section{Results from spectral fits}
\label{sec:res}

\subsection{NGC 3201: a showcase for the technique}
\label{sec:3201}

\begin{figure*}
	\includegraphics[width=0.95\textwidth]{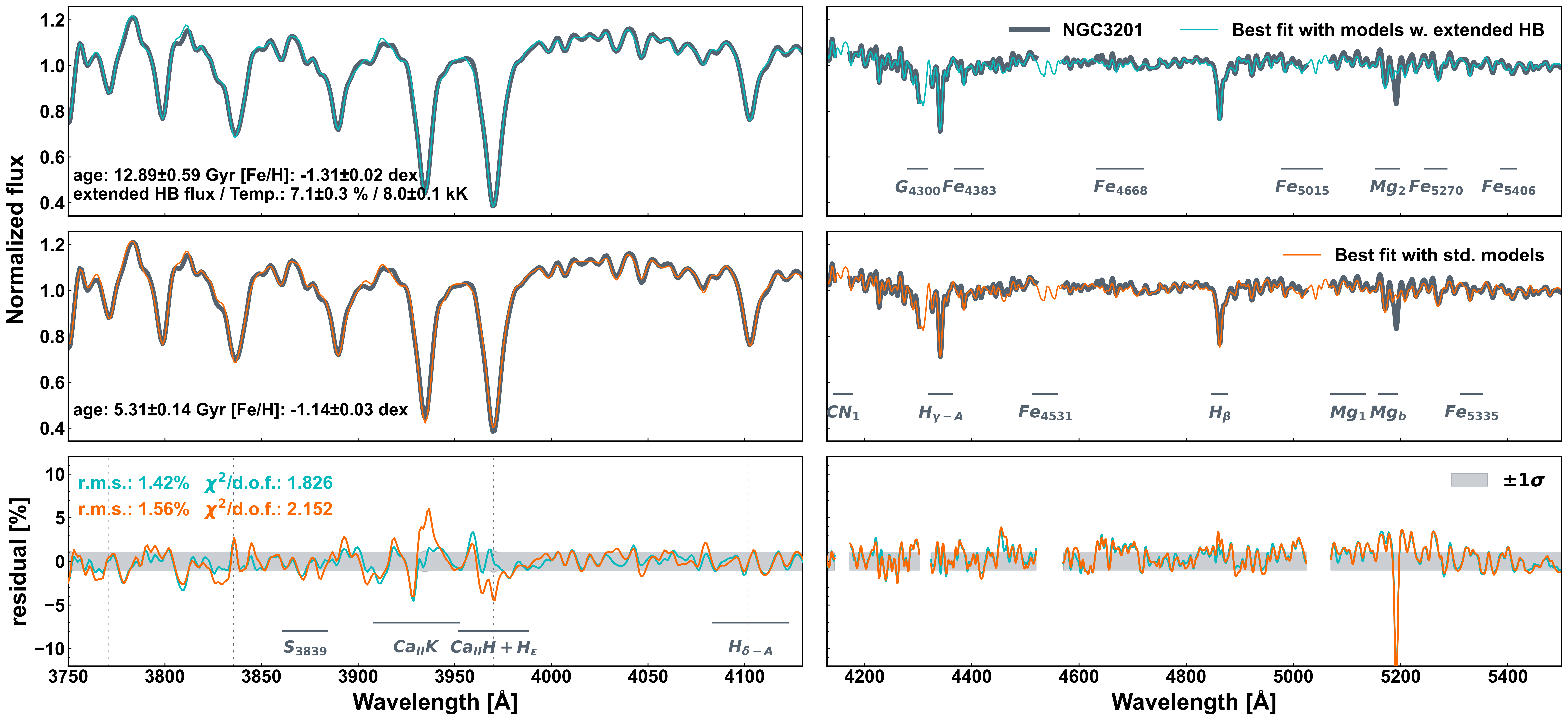}
    \caption{Fits to the optical integrated spectrum of NGC 3201. \textbf{Left:} Zoom on the spectral region around the \ion{Ca}{ii}\,K and \ion{Ca}{ii}\,H\,+\,H$_\epsilon$ features where the different models disagree the most. \textbf{Right:} Remainder of the fitted spectrum. Here the gaps in the observed spectrum correspond to the masked regions with corrupted data. \textbf{Top:} In grey we show the normalized spectrum of NGC 3201, and in blue the best fit solution using models with extended HB stars. 
    \textbf{Middle:} Like top but for standard models (orange line), i.e. without extended HB stars. \textbf{Bottom:} Residuals, (data-model)/model, of our best fit solutions. Vertical lines, mark the positions of Balmer lines in this wavelength range. For reference the $\pm 1 \sigma$ uncertainties of the observed spectrum are shown as shaded regions. The rms values of the residual spectra and reduced $\chi^2$ of the fits are quoted in the top right.}
    \label{fig:spec_good}
\end{figure*}

Before we present the results for the entire sample used in our analysis, we will focus on one of our targets to show a detailed example of the problem in question. In Fig. \ref{fig:spec_good} we show the results of our spectral fits to a target with an extended HB population, NGC 3201 (the spectra with the lowest SNR in our \citeauthor{Schiavon05} sample), using models with (top) and without (middle) an extended HB. From these two panels we can see that both solutions produced very similar spectra, yet the ages of the best fit models are very different.

In the bottom panel of Fig. \ref{fig:spec_good} we show the residuals, (data-model)/model, of our best fit models. In the bottom left panel we focus on the region where the different models diverge the most. Although overall both models reproduce well the observed spectrum, with differences between the data and models of the order of $\sim2\%$, the region around the \ion{Ca}{ii}\,K and \ion{Ca}{ii}\,H\,+\,H$_\epsilon$ features shows different behaviour between models. Specifically, we find that the \ion{Ca}{ii}\,K in the standard model solution (young) is significantly ($>3 \sigma$) stronger  than the observed feature, while the \ion{Ca}{ii}\,H\,+\,H$_\epsilon$ is weaker. In Section \ref{sec:spur} we will show that this behaviour is characteristic of solutions with spurious young ages, but first we will present the results obtained for our entire sample and how these compare with the reference age, metallicities and HB properties inferred with independent methods.

The best fit to the integrated spectrum of NGC 3201 using standard models is {$5.31 \pm0.14 {\rm~Gyr}$}. On the other hand the results of the spectral fits using models that include the contribution of the extended HB yields an age that is significantly older, {$12.89\pm 0.59 {\rm~Gyr}$}. 
 Although the extended HB population of NGC 3201 accounts for only a small fraction of the integrated light ({$\sim5\%$} of the flux in the $F606W$ band), accounting for its presence has a strong influence in our inferences from the integrated spectra, e.g. in this case a formation redshift of {$z<0.5$} versus { $z>6$}. 

 When we compare these results to previous studies, we find that when we include in our model spectra  the extended HB we find a better agreement with the ages inferred from fits to the CMD, e.g.: $10.24\pm 0.38 {\rm~Gyr}$ \citep{Forbes10}, $12.00\pm 0.75 {\rm~Gyr}$ \citep{Dotter10,Dotter11} and $11.5\pm 0.38 {\rm~Gyr}$ \citep{VandenBerg13}.

\begin{figure}
    \begin{center}
        	\includegraphics[width=0.45\textwidth]{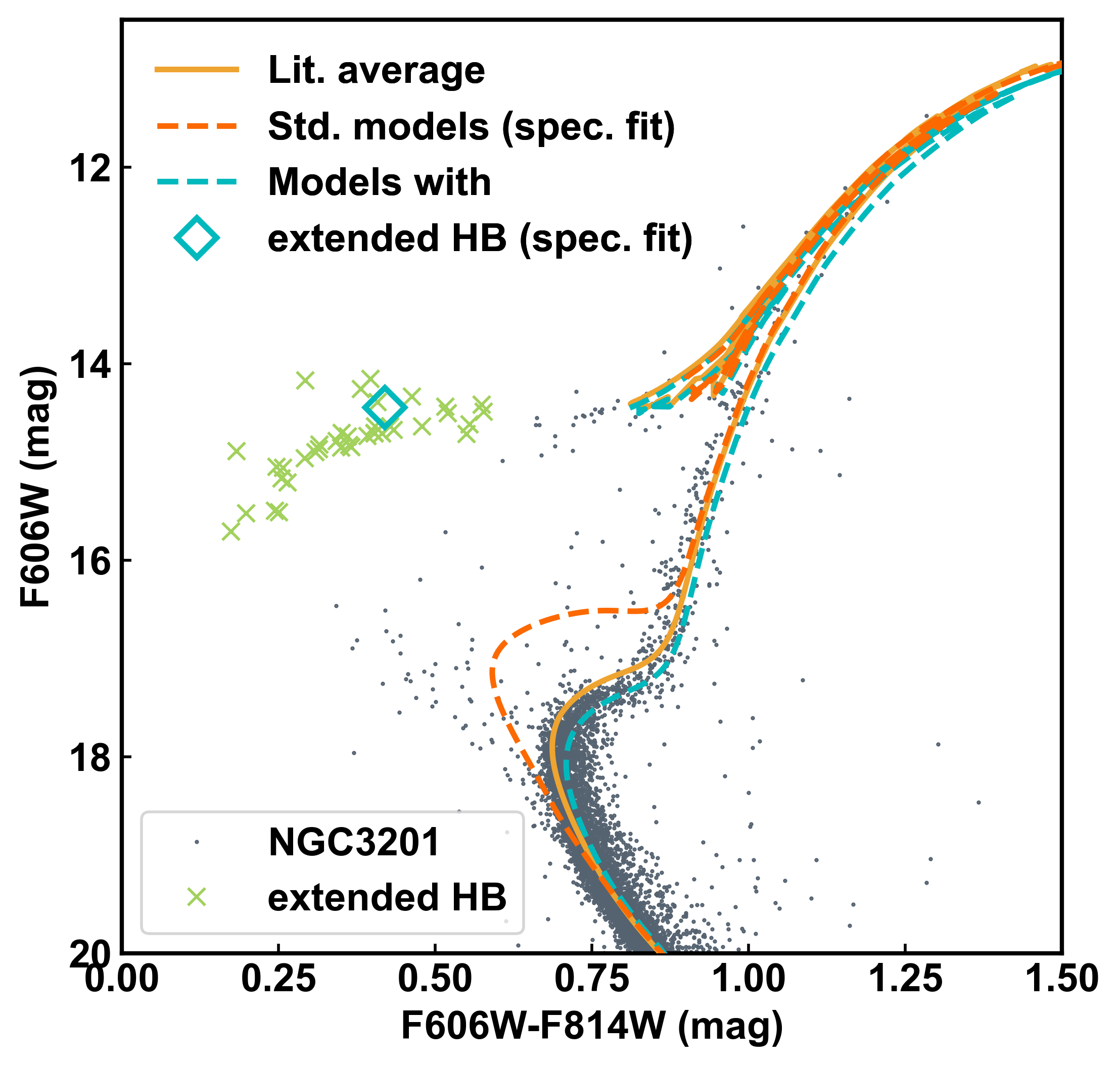}
    \caption{Optical CMD of NGC 3201, with its extended HB stars shown with green crosses. The yellow isochrone ($11.24 {\rm~Gyr}$, ${\rm [Fe/H]}=-1.4 {\rm ~dex}$) represents an average of literature results (see Table \ref{tab:lit}). The orange isochrone ($5.31 {\rm~Gyr}$, ${\rm [Fe/H]}=-1.14 {\rm ~dex}$) corresponds to the solution of our fit to the spectrum of NGC 3201 using standard models. The blue isochrone ($12.89 {\rm~Gyr}$, ${\rm [Fe/H]}=-1.31 {\rm ~dex}$) represents the solution of the spectral fit with models including the contribution of extended HB stars. The diamond represents the light weighted mean properties of the recovered BHB population inferred from the fit to the integrated spectrum. \emph{Note that none of these are actual fits to this CMD}. Here we adopted the same reddening and distance modulus for all isochrones, values reported in Table \ref{tab:lit}. 
 }
    \label{fig:cmd}
    \end{center}
\end{figure}

 In Fig. \ref{fig:cmd} we show the CMD of NGC 3201 and the isochrones corresponding to the solutions of our spectral fits (dashed lines). In orange we show the solutions obtained using standard models ({$5.31 {\rm~Gyr}$}, {${\rm [Fe/H]}=-1.14 {\rm ~dex}$}) and in blue the results of the spectral fit using models including extended HB stars ({$12.89 {\rm~Gyr}$}, { ${\rm [Fe/H]}=-1.31 {\rm ~dex}$}). The solid yellow line represents the reference value ($11.2 {\rm~Gyr}$, ${\rm [Fe/H]}=-1.4 {\rm ~dex}$) from the literature (see Section \ref{sec:phot_ages}). From this figure, we can reject the (young) solution of the fit with standard models at high confidence.

\subsection{Ages}

\begin{figure}
\begin{center}
	\includegraphics[width=0.45\textwidth]{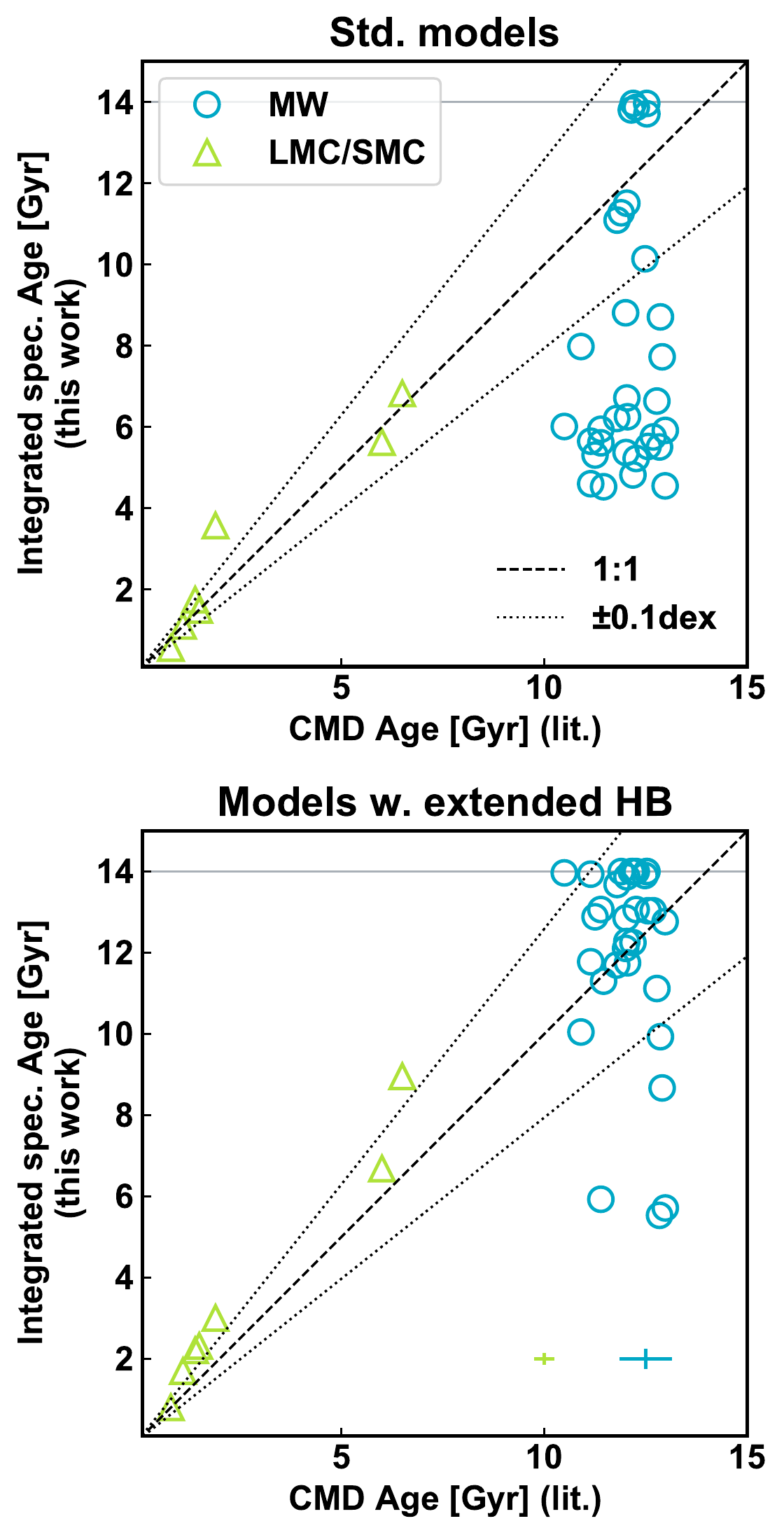}
    \caption{\textbf{Top:} Comparison of the literature ages of globular cluster in our sample (x-axis) vs. the ones derived from fits to their integrated light using standard stellar population models that, i.e. not including blue horizontal branch stars (y-axis). The square and triangle symbols represent the clusters from the MW and LMC/SMC respectively. \textbf{Bottom:} Similar to top, but for models including extended horizontal branch stars. The error bars show the averaged statistical uncertainties of our measurements and standard deviation of the literature values. The grey horizontal line represents the upper limit of our age prior.}
    \label{fig:ages}
\end{center}
\end{figure}

Figure \ref{fig:ages} shows a comparison of the ages obtained from the fits to the integrated spectra and the values obtained from CMD analyses. The top panel presents the results using standard stellar populations models, i.e. without including the contribution of an extended HB population (values presented in Table \ref{tab2}).
The bias towards younger ages found in old ($>9 {\rm ~Gyr}$) clusters when using this type of models is clear, and is consistent with what has been found in other recent works \citep[e.g.][]{Conroy18,Goncalves20,Johnson20}.

From these experiments using standard stellar population models we find that just 9/32 of the old targets have ages within $0.1 {\rm ~dex}$ of their CMD reference value, 
while most them (23/32) have an age underestimated by $0.2-0.4 {\rm ~dex}$. On the other hand, the ages of the younger targets ($<9 {\rm ~Gyr}$, triangles) in our sample are in good agreement with their literature value, typically within $\lesssim 1 {\rm ~Gyr}$.

In the bottom panel of Fig. \ref{fig:ages} we show the ages obtained after including a contribution of extended HB stars in the models (values presented in Table \ref{tab:lit}). The bias towards young ages has been reduced significantly. Although the statistical uncertainties of our measurements are virtually the same for both cases (few hundred Myr, shown as error bars), now the agreement with the literature (CMD) values is much better, with the ages derived from the integrated spectra for 26/32 of the old targets within $0.1 {\rm ~dex}$ of the CMD values (we take this value as the systematic error of our measurements).
That said, we find that a small fraction (3/32) our fits did not converge to an extended HB solution that would improve the results in the age recovery, namely: NGC 2298, NGC 5946, and NGC 7078 (circles with ages from integrated spectra $\sim6 {\rm ~Gyr}$). For this targets the ages from the integrated light analysis are still underestimated by {$\sim0.3 {\rm ~dex}$}. In Section \ref{sec:hb}, we will discuss this in more detail.

These experiments also show that the age of the young targets is overall recovered well from the integrated spectra (in most cases within a few hundred Myr of the CMD ages), even when the fits that included an extended HB\footnote{We note that stars that burn He in their core at these ages (1-6 Gyr) are not usually referred to as `HB stars', but in order to simplify our terminology/discussions we will refer to them as such.}.
That said, we remind the reader that targets of such ages do not manifest extended HBs, so modelling of the extended HB component is not recommended for targets $<9 {\rm ~Gyr}$ (more on this in Section \ref{sec:fut}).

\subsection{Metallicities}

\begin{figure}
\begin{center}
\includegraphics[width=0.45\textwidth]{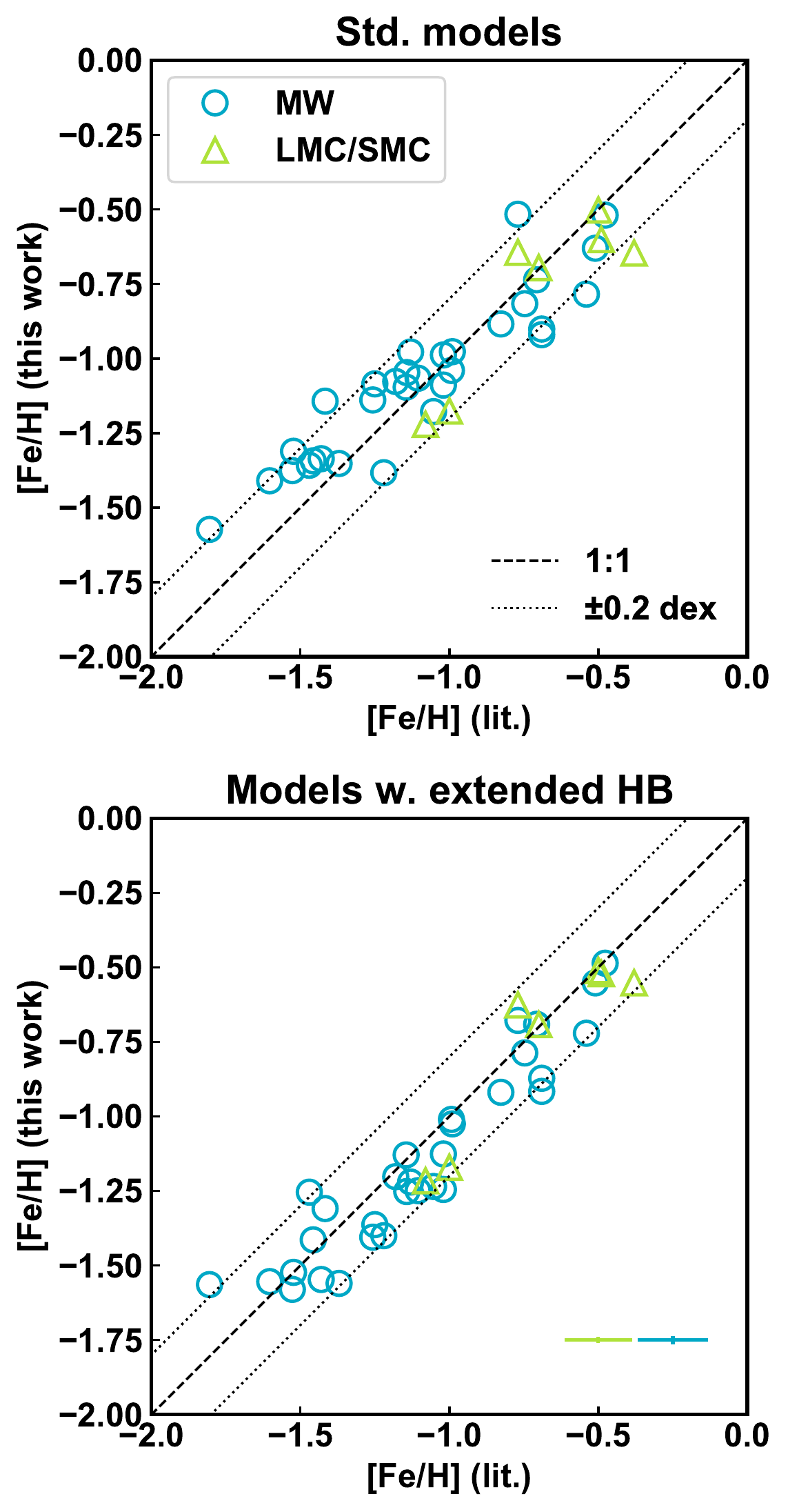}
    \caption{Similar to Fig. \ref{fig:ages} but for [Fe/H].}
    \label{fig:fe}
\end{center}
\end{figure}

In general, the [Fe/H] inferred from the fits to the integrated light are in good agreement ($<0.2 {\rm ~dex}$ systematic errors) with the literature values, mostly obtained using high resolution spectra of individual stars, see Fig. \ref{fig:fe}. Like in Fig. \ref{fig:ages}, here the vertical error bars represent the formal statistical uncertainties. When using standard models, we find that the residual distribution is more symmetrical than when using models including extended HB stars. However, these systematic differences are not significant as they are still within the typical uncertainties in the literature $\sim0.1 {\rm ~dex}$.

The agreement between the metallicity inferred using models with and without extended HB stars is not a surprise, since the atmospheres of extended HB stars are charaterized by weak metal lines product of their hot temperatures (e.g. comparison between \ion{Ca}{ii} and H lines in left panel of Fig. \ref{fig:ew}). Thus the presence of these stars in the integrated light of a stellar population have little influence on features sensitive to metallicity. Along these lines we note that there does not seem to be an age-metallicity degeneracy when inferring spurious young solutions, even though the age could be underestimated by a factor of $\sim2$ in a few cases.

\subsection{Extended HB parameters}
\label{sec:hb}

\begin{figure}
\begin{center}
	\includegraphics[width=\columnwidth]{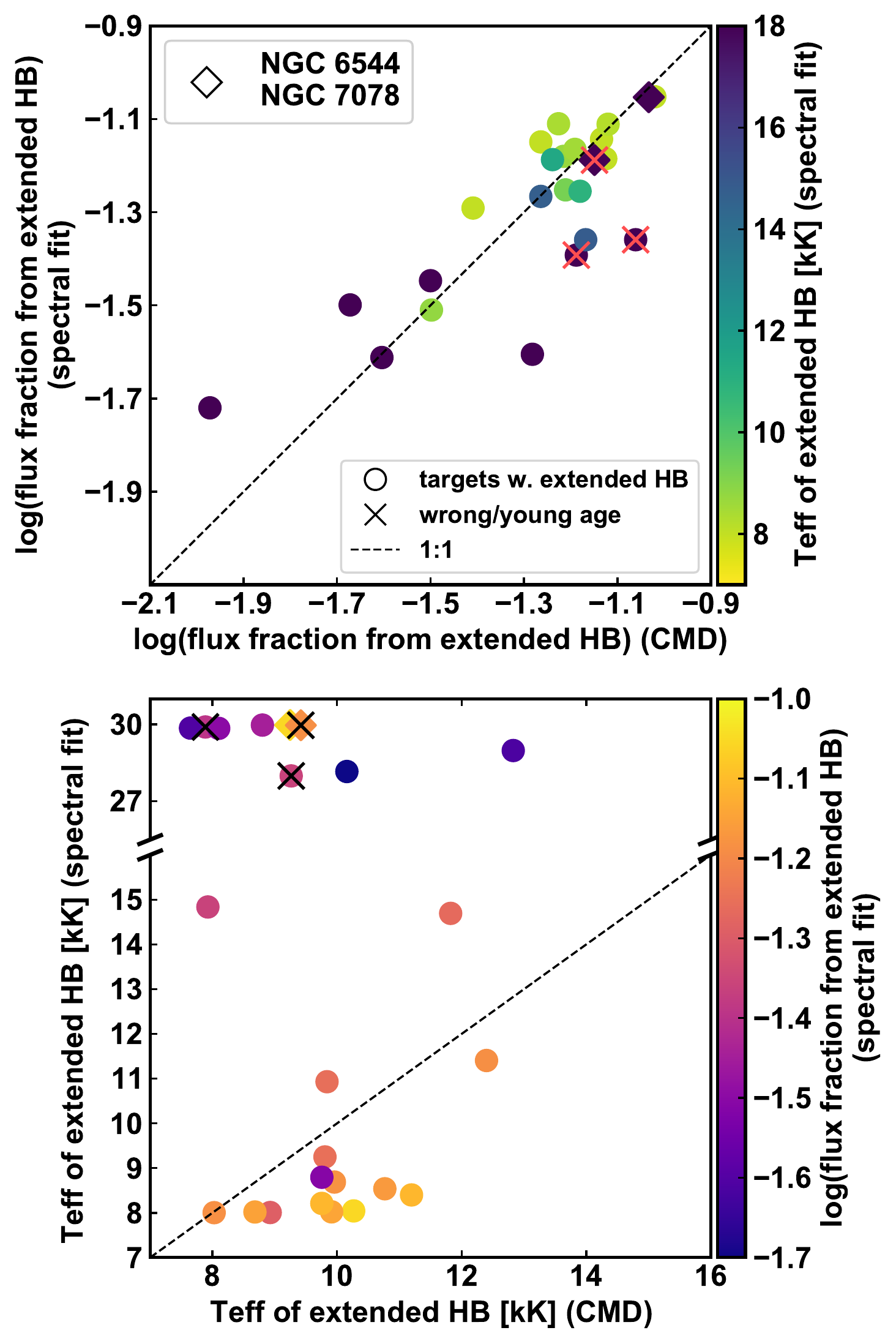}
    \caption{Comparison between the extended HB branch parameters derived from the CMD and integrated spectra. \textbf{Top:} Fraction of the flux of the extended HB component with respect to the integrated light. On the horizontal/vertical axis we show the value derived form the CMDs/spectral fits, respectively. For targets with WFPC2 CMDs we compare the flux in the $F555W$ band, while for clusters with ACS CMDs we use the $F606W$ band. Colour coded is the inferred [Fe/H] from our fits. 
    The crosses mark the clusters with extended HB for which our spectral fits recovered a young age. \textbf{Bottom:} On the x-axis we show the light weighted mean of the $T_{\rm eff}$ of the extended HB population derived from the CMD on the y-axis the flux of the extended HB component inferred from our spectral fits. Note that the y-axis of this figure breaks at {$T_{\rm eff}=16{\rm ~kK}$} and resumes again at $T_{\rm eff}=25.5{\rm ~kK}$ in order to maintain clarity while including the large outliers. The colours show the  flux contribution of the extended HB inferred from our spectral fits. NGC 6544 and NGC 7078 are shown with diamonds.}
    \label{fig:hb}
\end{center}
\end{figure}

In the top panel of Fig. \ref{fig:hb} we compare the values of the flux contribution of the extended HB population to the integrated light of the cluster inferred from the spectral fit with the values derived from the CMD. As mentioned in Section \ref{sec:phot_hb}, the photometry of the clusters analysed here is either from the WFPC2 or the ACS Galactic globular cluster surveys. Although these surveys do not have filters in common, we are able to calculate the flux contribution of extended HB population inferred from our spectral fits in both sets of filters. For the targets with WFPC2 photometry we use the flux in the $F555W$ band while for the ACS targets we compare the flux in the $F606W$ band.

The contribution to the integrated light of the extended HB population of our targets is very diverse, covering one order of magnitude between extremes (see Fig. \ref{fig:hb}). On one end we have NGC 6388 where the extended HB population represents only $\sim1 \%$ of the integrated light, while for a target like NGC 1904, this can be as much as $\sim 10\%$. From our spectral fits we were able to recover this parameter and found consistent values between the CMD analysis and our results with differences $<0.15 {\rm ~dex}$ for 18/24 of the targets in our sample (see Fig. \ref{fig:hb}). However, the (light weighted) mean temperature of the extended HB population proved to be more difficult to infer from our spectral fits (bottom panel).

For most (14/24) of our targets we are able to recover the mean temperature of the extended HB population within { $<30\%$} from their reference value. However, the posterior temperature distribution was unconstrained (reached the limits of the prior) for {9/24} targets with extended HB stars. Interestingly, even though the temperature of the extended HB population inferred from the spectra of NGC 2808, NGC 6171, NGC 6362, NGC 6388, NGC 6544 and NGC 6638 was significantly higher than ($>15 {\rm ~kK}$) the value inferred from the CMD, we were still able to improve the age by including a population of extended HB. This behaviour and other results will be discussed in more detail in Section \ref{sec:disc}.

\subsection{Comparison with previous work}

\begin{figure}
	\includegraphics[width=\columnwidth]{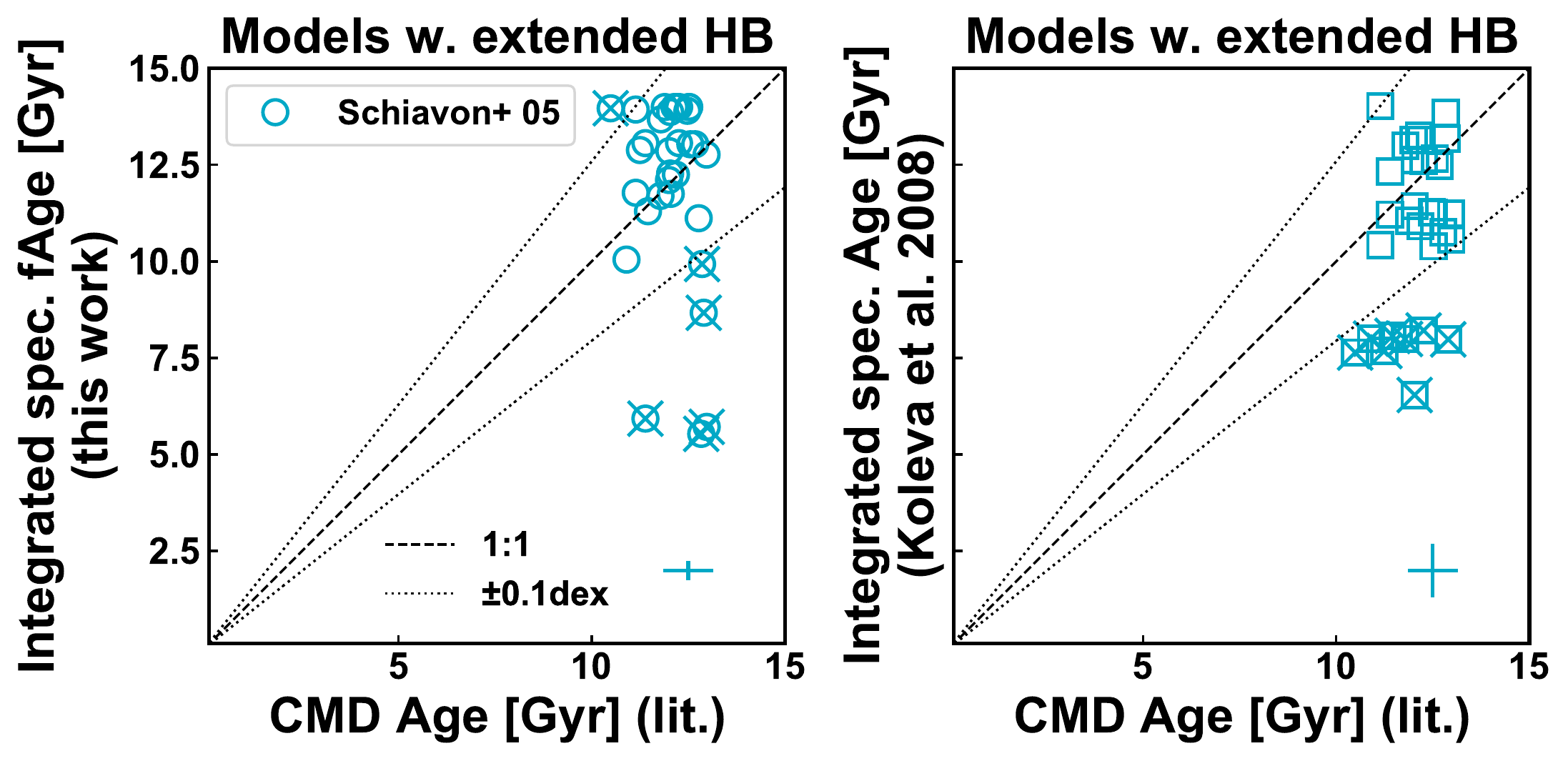}
    \caption{Comparison of the ages inferred for the \protect \cite{Schiavon05} spectra by two groups that model the contribution of extended HB stars during the spectral fit. In the left panel we show the results from this study (similar to right panel of Fig. \ref{fig:ages}) and in the right panel we show the results reported by \protect \cite{Koleva08}. Targets for which modelling the extended HB produced age solutions $>0.1 {\rm ~dex}$ away from the reference value are shown with crosses.}
    \label{fig:ages_koleva}
\end{figure}

As mentioned in Section \ref{sec:intro}, our approach to modelling the contribution of the extended HB stars to the integrated spectra of stellar populations is similar to the one implemented in \cite{Koleva08}. In Fig. \ref{fig:ages_koleva} we carry out a comparison of the ages derived by both studies to the same control sample \citep[the spectra from ][]{Schiavon05}.
Overall the results from both studies are comparable with similar systematics with respect to the literature value ($\sim0.1 \mbox{~dex}$), with only {six} (in this work) and eight \citep[in][]{Koleva08} targets beyond this threshold. Unfortunately the data necessary for a comparison of the parameters recovered for the extended HB is not available.

\section{Discussion}
\label{sec:disc}

The results presented in the previous Section show that the age and flux of the extended HB population inferred from our spectral fits are overall accurate when compared to the values inferred from the CMDs. Similarly, the metallicity inferred from the integrated spectra is consistent with the one obtained using high resolution spectra of individual stars. All this leads us to conclude that our approach to the modelling of the integrated spectra of stellar populations produces comparable results in the recovery of these parameters than the ones obtained with more (observationally) expensive data sets that will not be available for distant targets.

In this section we discuss in more detail some of these results and highlight a few aspects where there is room for improvement.

\subsection{Spurious young solutions}
\label{sec:spur}

There are a few cases where for some targets with old CMD ages, the spectral fits including an extended HB population still produced a young ($\sim 6 {\rm ~Gyr}$, i.e. younger by $\sim0.3 {\rm ~dex}$) solution, namely NGC 2298, NGC 5946, and NGC 7078. The potential of this technique for the study of unresolved stellar populations relies on being able to identify from the spectral fits when a young solution is genuine and when it is spurious.

\begin{figure*}
	\includegraphics[width=\textwidth]{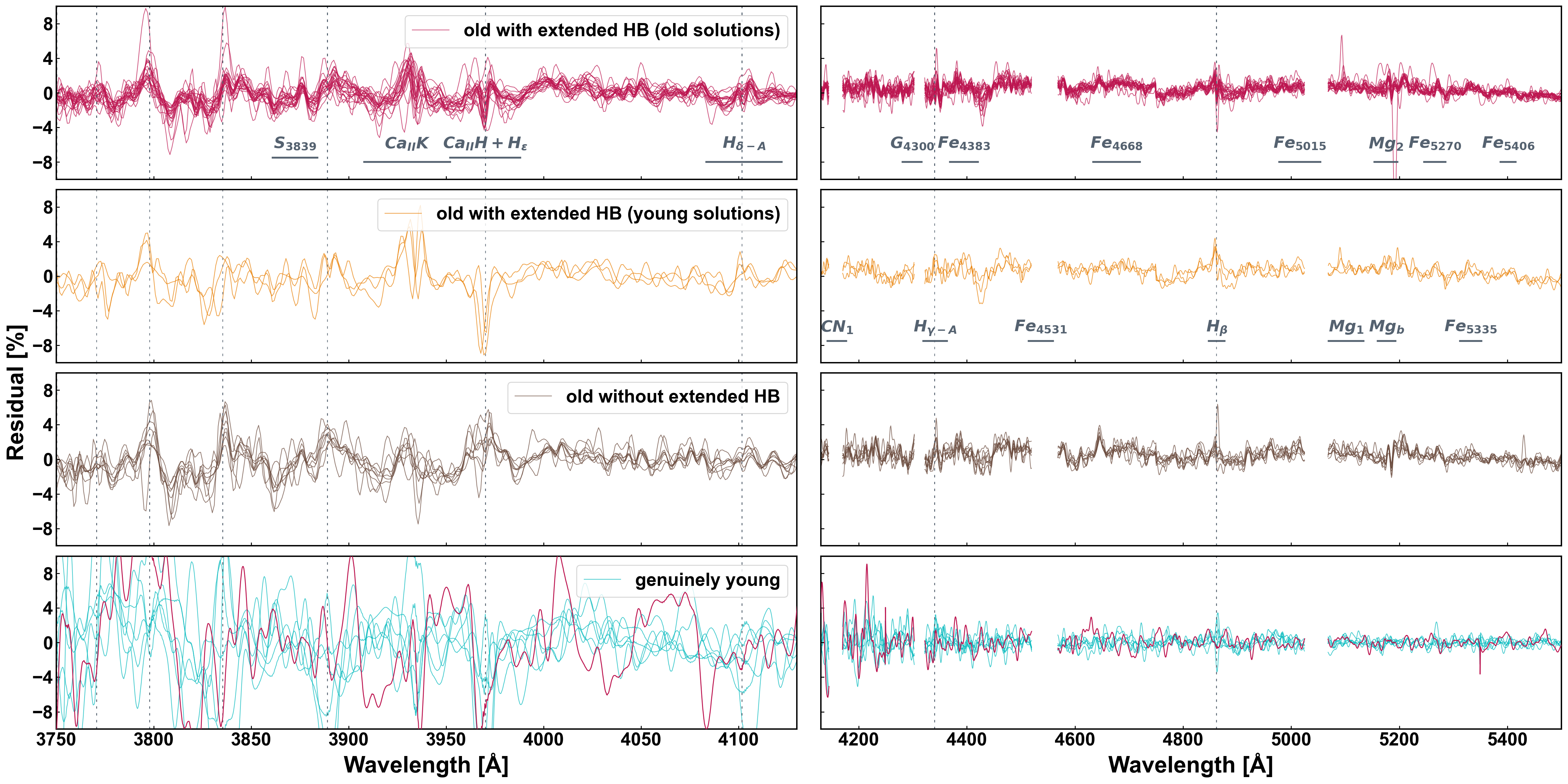}
    \caption{Residuals, (data - model) / model, of the spectral fits with models including extended HB stars around \ion{Ca}{ii}\,K and \ion{Ca}{ii}\,H\,+\,H$_\epsilon$ (left) and the rest of the fitted spectrum (right). The first two rows show the residuals for the targets with old CMD ages that host an extended HB population. The top one, representing the cases where the spectral fits recovered an old age while the second one shows the cases where the fits yielded a young (incorrect) solution. The third row shows the residuals for targets with old CMD ages that do not host an extended HB population, and the bottom row shows the residuals for the targets with young CMD ages. In the bottom panel, the red line represents Kron 3. As mentioned in Section \ref{sec:data}, the spectra of the young targets is noisier than for the older ones, hence the noisier residuals. The distinct signatures in second row on the left allow us to recover when things go wrong in the age determination of old clusters with extended HB. These are different from the residuals found for genuinely young targets. The position of Balmer lines are shown with dashed lines.}
    \label{fig:ehb_residuals}
\end{figure*}

Our experiments suggest that this is possible by analysing the residuals of our spectral fits. In Fig. \ref{fig:ehb_residuals}, we show the residual spectra, (data - model) / model, of the fits including an extended HB component for different sub-samples of the targets analysed here. On the top two panels, we show the residual of the targets with old ages and an extended HB according to their CMDs. The first one shows the results for the targets for which we were able to recover successfully an old ($\gtrsim9 {\rm ~Gyr}$) solution, while the second one shows the residuals of the five targets where the spectral fits inferred ages significantly ($\sim0.3 {\rm ~dex}$) younger than the CMD values. Similarly the third and fourth panel presents the residuals of the spectral fits to the sub-sample of targets with old CMD ages without an extended HB population and targets with young CMD ages, respectively. As in Fig. \ref{fig:spec_good}, the left panels show a zoom on the region where the best fit models display the largest differences with respect to the observed spectrum while the right panels show the rest of the fitted spectral range.

Although in all cases the fits to the spectra are very good --with differences between the observed and model spectra of the order of few percent-- there are some systematic trends in these residuals. For example, when the fits produced (correct) old solutions, i.e. rows one and three, the differences between observed and model spectra are very similar.
In both cases, the Balmer lines (dashed lines) are comparable or slightly stronger in the best fit model spectra including an extended HB contribution than in the data (positive residuals). On the other hand, for genuinely young targets (bottom panel of Fig. \ref{fig:ehb_residuals}), the best fit models with an extended HB population produced spectra with weaker Balmer lines than the ones observed in the spectra of young targets (negative residuals).

Interestingly, the residual spectra obtained when the spectral fits assigned a young solution to an old target with an extended HB (second row), are very distinct to the rest. In particular, the behaviour of the \ion{Ca}{ii}\,K and \ion{Ca}{ii}\,H\,+\,H$_\epsilon$ stand out. As mentioned in Section \ref{sec:3201}, our models find difficult to reproduce the observed \ion{Ca}{ii}\,K and \ion{Ca}{ii}\,H\,+H$_\epsilon$ features making it one the worst fit regions in the spectra (with difference of up to $\sim8\%$ in the core of these features). For the spurious young solutions we find that the \ion{Ca}{ii}\,K is systematically stronger in the best fit spectra with extended HB while the \ion{Ca}{ii}\,H\,+\,H$_\epsilon$ is systematically weaker.

\begin{figure}
\begin{center}
	\includegraphics[width=0.45\textwidth]{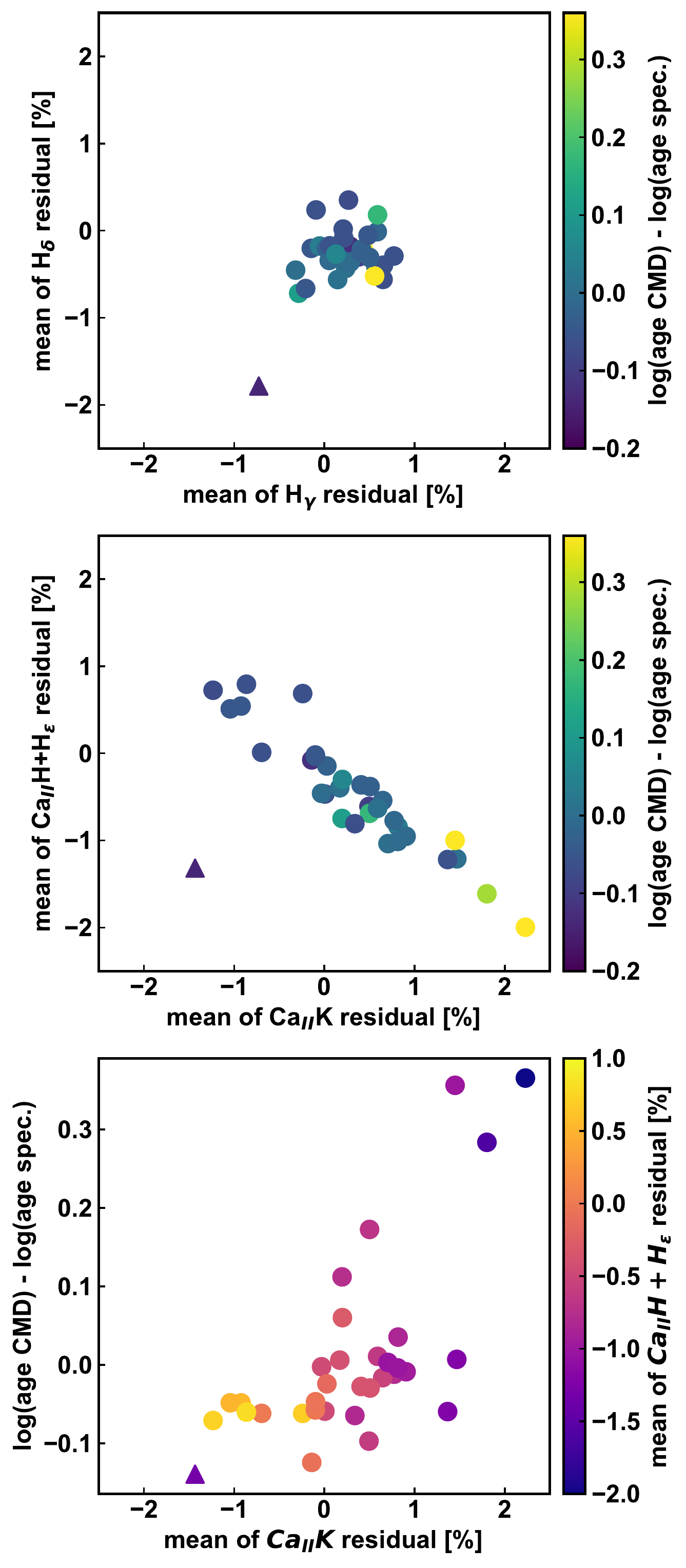}
    \caption{\textbf{Top:} Strength of the residuals of the spectral fits around H$_\gamma$ and H$_\delta$ for the Milky Way targets shown in Fig. \ref{fig:ehb_residuals} (see text). Positive (negative) values correspond to targets where the best fit model has a stronger (weaker) features than the data. Colour coded is the difference in age between the reference age and the age inferred from the spectral fits using models with extended HB stars. \textbf{Middle:} Similar to the top panel but for the \ion{Ca}{ii}\,K and \ion{Ca}{ii}\,H\,+\,H$_\epsilon$ features. Targets with underestimated ages tend to be in the lower right part of this diagram, while targets with overestimated ages tend to lie towards the upper left. \textbf{Bottom:} Similar to middle but now showing the age difference as a function of \ion{Ca}{ii}\,K residual strength (colour coded by \ion{Ca}{ii}\,H\,+\,H$_\epsilon$ residual strength). When the models overestimate the strength of \ion{Ca}{ii}\,K the inferred age is systematically underestimated (\ion{Ca}{ii}\,K residuals $\sim2\%$, age difference $\sim0.3 {\rm ~dex}$). Kron 3 is shown as a triangle.}
    \label{fig:ca_res}    
\end{center}
\end{figure}

To better illustrate this trend and its origin, we quantify strength of the residuals by measuring their mean value around different spectral features and present the results in Fig. \ref{fig:ca_res}. For reference, in the top panel we show the behaviour of the residuals around two strong Balmer lines, H$_\gamma$ and H$_\delta$. We colour coded the difference between the reference age and the age inferred from the spectral fits. 
Here we show that our models do a good job reproducing the observed Balmer lines, with very small ($<1\%$) residuals distributed around zero regardless if we inferred an accurate (old) age or not.

In the middle panel of Fig. \ref{fig:ca_res} we show the strength of the \ion{Ca}{ii}\,K and \ion{Ca}{ii}\,H\,+H$_\epsilon$ residuals. Here a clear (negative) correlation is present, reflecting that the intrinsic strength of the \ion{Ca}{ii}\,K and \ion{Ca}{ii}\,H\,+H$_\epsilon$ are not entirely independent as shown in Fig. \ref{fig:ew}.

Briefly, Fig. \ref{fig:ew} shows that \ion{Ca}{ii}\,K is sensitive to the age of the stellar population. On the other hand \ion{Ca}{ii}\,H\,+H$_\epsilon$ is a blend of two features: \ion{Ca}{ii}\,H which is sensitive to age, and H$_\epsilon$ which is as a first order sensitive to the temperature of the HB and in second order sensitive to age.

From the top two panels of Fig. \ref{fig:ca_res} we find that, having fit the Balmer lines (either driven by a hot HB or by turn off stars, see Section \ref{sec:hb_disc}), the relative strength between the observed and modelled \ion{Ca}{ii}\,K (and \ion{Ca}{ii}\,H\,+H$_\epsilon$) encodes information about how close the model age is to the observed spectrum. If the observed \ion{Ca}{ii}\,K is weaker than the best fit model, i.e. large ($\sim2\%$) residuals, this would suggest that the age of the best fit model is likely to be underestimated (see bottom panel of Fig. \ref{fig:ca_res})\footnote{We found the same behaviour when analysing the residuals of the spectral fits with standard models (cf. Fig. \ref{fig:ehb_residuals2})}. We propose to use this behaviour to identify spurious solutions.

To summarise, Figs. \ref{fig:ehb_residuals} and \ref{fig:ca_res} show that \emph{the residual of our the spectral fits can be used to distinguish between genuine and spurious solutions}. Specifically, spurious young solutions are characterised by models where the \ion{Ca}{ii}\,K line is stronger than the observed one and the model \ion{Ca}{ii}\,H\,+\,H$_\epsilon$ is weaker than the observed feature.

That said, the systematic differences between the observed spectra and our best fit models shown in Fig. \ref{fig:ehb_residuals} and \ref{fig:ca_res}, suggest that there is still room for improvement in our models. At the moment our modelling of the extended HB population is very simple/idealized, as it is represented by the spectrum of a single hot HB star, however in a future work will explore a more realistic (multi-component) modelling of the extended HB star population.

\subsection{Spurious old solutions}

In the bottom panel of Figure \ref{fig:ages}, we find that for one of our young ($\sim7 {\rm ~Gyr}$) SMC cluster, Kron 3,  our fit that includes the effect of extended HB stars yields an age that is slightly ({ $\gtrsim0.13  {\rm ~dex}$}) older than the reference value. As mentioned in Section \ref{sec:data}, the spectra of the young clusters has a significantly lower SNR than the older ones (note large residuals in Fig. \ref{fig:ca_res}) and have been included in this analysis just for reference. That said, for Kron 3 the fit with standard stellar population models reproduced slightly better \ion{Ca}{ii}\,K and Balmer lines like ${\rm H}_\gamma$ and ${\rm H}_\delta$ than the model with an extended HB converged to a solution consistent with the reference value.

\subsection{Recovery of HB parameters}
\label{sec:hb_disc}

As mentioned in Section \ref{sec:hb}, in the bottom panel of Fig. \ref{fig:hb} we find nine targets where the inferred temperature of the extended HB population is significantly hotter than the reference value obtained from the CMD. For ({3/9}) of these targets our spectral fits did not converge on a solution consistent with the reference value (i.e. young solutions, shown with crosses).

\begin{figure}
	\includegraphics[width=\columnwidth]{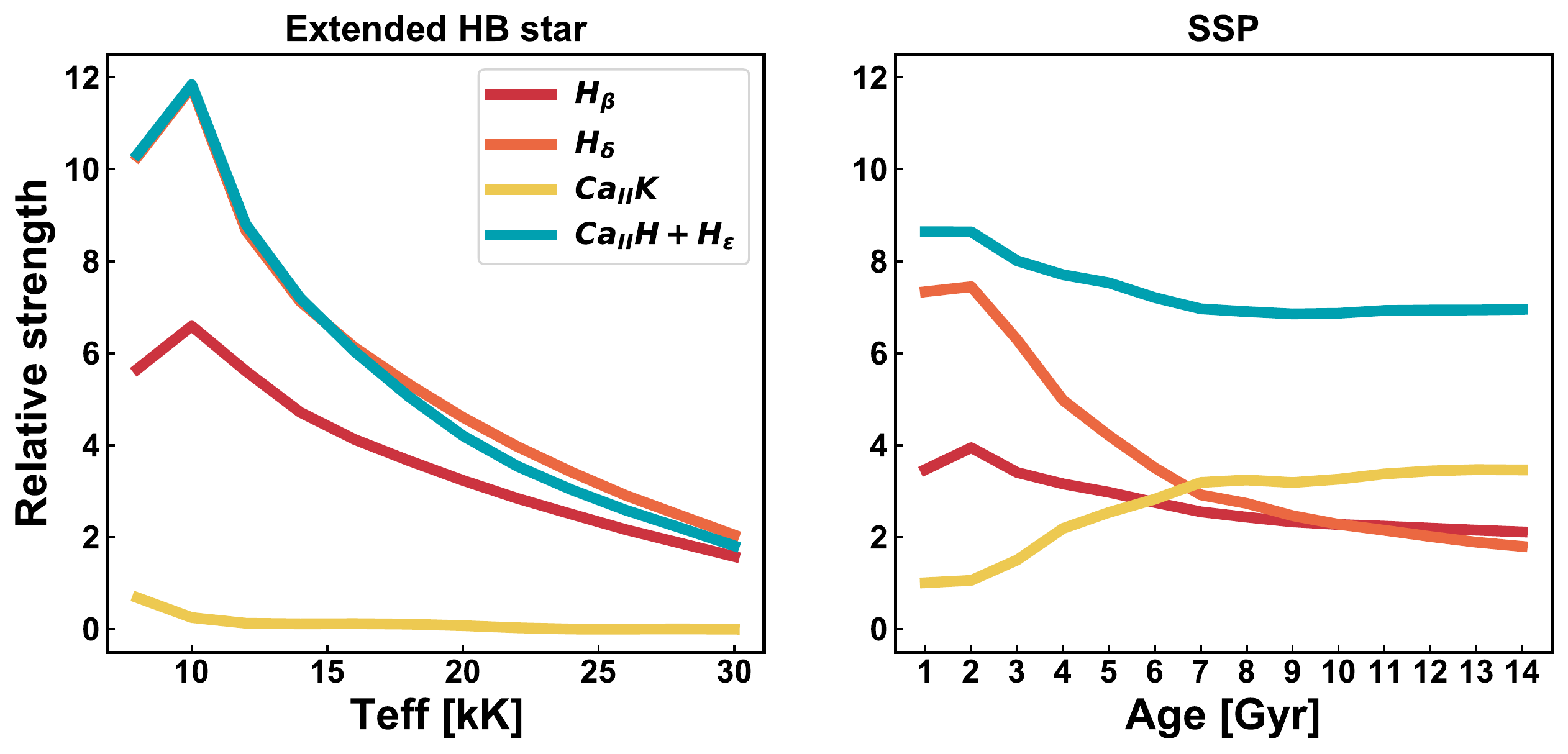}
    \caption{\textbf{Left:} Strength of some H and \ion{Ca}{ii} spectral features as a function of temperature for extended HB stars. \textbf{Right:} Similar to left but for a standard simple stellar population model as a function of age.}
    \label{fig:ew}
\end{figure}

We can understand this behaviour by looking at the strength of some key spectral features and their sensitivity to age and temperature. In the left panel of Fig. \ref{fig:ew} we show how the strength of some Balmer and \ion{Ca}{ii} lines in extended HB stars (with ${\rm [Fe/H]}=-1.5 {\rm ~dex}$) change as a function of temperature. On the right, we show the evolution of these indices as a function of age for standard simple stellar population models of the same metallicity.

Similar to what is found when using standard stellar population models, in these three cases the age of the cluster is the parameter driving the strength of the Balmer features, since the extended HB solutions for these particular clusters (i.e. very hot temperatures) produce spectra with very weak Balmer lines, see Fig. \ref{fig:ew}. 

We note that we find similar results for these targets after reducing the temperature for the higher limit of the prior for the extended HB stars (e.g. $15 {\rm ~kK}$ instead of $30 {\rm ~kK}$). In other words, these targets converged to a (mean) temperature for the extended HB population similar to the maximum temperature allowed by our prior.

In Fig. \ref{fig:hb} we also find six targets (NGC 2808, NGC 6171, NGC 6362, NGC 6388, NGC 6544 and NGC 6638) where the temperatures for the HB population inferred from our spectral fits is $>15 {\rm ~kK}$ hotter than their reference value, yet the spectroscopic ages of these targets are within $<0.1 {\rm ~dex}$ from the reference values ({$<0.17 {\rm ~dex}$} for NGC 6171). 

In the rest of this section we will focus on two clusters NGC 6544 and NGC 7078 to explain the disagreement between the temperature of the extended HB population inferred from the spectral fit and the one derived from the targets CMDs. For the former, we infer the right age while the spectral fit of the later yielded an old age. We chose these two targets as their extended HB population have a similar contribution to the integrated light and a light weighted mean temperature (i.e. $\log(f_{\rm hot})\sim-1{\rm ~dex}$ and $T_{\rm eff}\sim9{\rm ~kK}$ according to their CMDs). These targets are shown in Fig. \ref{fig:hb} with diamonds.

In Fig. \ref{fig:n7078} and Fig. \ref{fig:n6544} we compare the observed spectra of these clusters (grey lines) with the model spectrum of a population with the age and HB properties inferred from their CMDs with (red) and without (yellow) an extended HB population as well as the solution from our spectral fits (blue).

Here we see that for NGC 7078 the stellar population model given by the MIST isochrone with the age equal to the reference literature value does not reproduce the \citeauthor{Schiavon05} spectrum around ${\rm H}_\beta$, ${\rm H}_\gamma$ and ${\rm H}_\delta$ (shallow lines). However, when we add the spectrum of an extended HB component (as inferred from the CMD), we find a better agreement in these Balmer features (red spectrum). We also find an improvement in the \ion{Ca}{ii}\,K line. Having said that, the \citeauthor{Schiavon05} spectrum of NGC 7078 is best reproduced by the \texttt{alf} solution (blue).

On the other hand, for NGC 6544 we find that the standard stellar population model given by the MIST isochrone with the reference age does a good job reproducing ${\rm H}_\beta$, ${\rm H}_\gamma$ and ${\rm H}_\delta$ from the \citeauthor{Schiavon05} spectrum. The inclusion of an extended HB component with the properties inferred from the CMD, reduces the disagreement in the \ion{Ca}{ii}\,K line with respect to the model without the extended HB population, however, this is at the expense of producing slightly stronger Balmer features than the ones found in the \citeauthor{Schiavon05} spectrum. As with NGC 7078, the \texttt{alf} solution produces the best fit to the observed data.

We reach the same conclusions when we compare the observed spectra of these clusters with a model spectra of built using a linear combination of HB stars between 8 to 30 kK with the relative weights taken from the observed CMD. We found that the spectrum built using linear combination of HB stars still differed significantly from the observed spectrum of these targets. Such difference suggest that even a more sophisticated model of the HB population (i.e. including the spectra of more than one HB star) is not sufficient to reproduce the observed data.

We explored as well carrying out our fits between 4000-5000 \AA, i.e. excluding the \ion{Ca}{ii} and the higher order Balmer lines. From these experiments we found that inferred ages, metallicity and HB properties were in agreement (within the uncertainties) with the values obtained when fitting the extended wavelenght range (3750-5500 \AA). These experiments suggest that whatever is causing the models to converge to these particularly hot solutions is still dominant in the red end of the spectrum.

To summarize, even though our spectral fit solutions has HB properties (and sometimes ages) different to the reference values inferred from the analysis of the CMD, they do a better job at reproducing the observed spectra than the spectral models built with the properties inferred from the CMDs. In other words, in these cases the observed spectra does not look like what is expected according to their CMDs. Some possibilities for the disagreement between the models and observed data could be that the observed spectra are affected by foreground/background contamination, or some fundamental problem in our stellar population models, e.g. issues in some particular region in $\log({\rm g})$/$T_{\rm eff}$/[Fe/H] in our stellar library. Another possible explanation for this disagreement is stochasticity preventing a representative sampling of all stellar evolutionary phases (in particular the full population of rare stars like HB)  within the small aperture where the spectra was extracted \citep[see e.g.][]{Usher17,Usher19}\footnote{The spectra from \citeauthor{Schiavon05} were extracted using apertures of a few core radii. For reference, the core radius of a cluster in our sample hosts $<4\%$ of the stellar mass -- typically of the order of $10^3 {\rm ~M}_\odot$ \citep[see][]{Baumgardt19}}. 

We find similar conclusions for the other targets with hotter extended HB than the ones inferred from the CMDs. We note that for massive and distant globular clusters stochasticity is unlikely to be a major source of uncertainty, as their compact nature makes it easier to cover the entire extent of the cluster within the spectrograph's aperture. However, if a globular cluster has a mass sufficiently low such that not all evolutionary stages all well represented, the integrated spectra will be intrinsically affected by stochasticity -- regardless of the aperture of the spectrograph.

\section{Future steps}
\label{sec:fut}

As mentioned above, our modelling of the extended HB is very simple. We are just adding an extended HB contribution on top of the spectrum predicted from the standard isochrone. Formally this would tend to overestimate the flux contribution of this component, however, these effects seem to be smaller than our current systematic uncertainties (i.e. this is not evident in top panel of Fig. \ref{fig:hb}). A future iteration of our code, should aim at redistributing the temperature of HB rather than adding the flux of extra stars, as it is implemented in other stellar population models \citep[e.g.][]{Lee00,Maraston05,Conroy09,Conroy10,Chung17}.

The SED information from standard optical broad-band photometry could also be included in future analyses in order to break the degeneracy between age and HB properties (as long as the degeneracy with the reddening is not too severe). Although the normalised spectrum of a young population and an old one with extended HB stars are very similar (see Fig. \ref{fig:spec_good}), their SED should be easier to distinguish, specifically in the UV.

Finally, in our experiments we found that the reduced $\chi^2$ is systematically lower for the fits that include an extended HB component (see Table \ref{tab:lit} and Table \ref{tab2}). This means that we need a different criterion to determine if the modelling of the extended HB is necessary/appropriate or not.
Besides the approaches mentioned in the introduction \citep[e.g.][]{Schiavon04,Percival11}, another way to do this could be to introduce a prior that allows solution with extended HB contributions only for old (e.g. $>9 {\rm ~Gyr}$) ages, avoiding inappropriate/nonphysical young solutions with extended HBs. Similarly for targets where our fits with standard models yield an old solution (i.e. clusters with no extended HB) we could keep that solution instead of fitting an unnecessary extended HB component.

\section{Summary}
\label{sec:sum}

The determination of age of stellar populations from their integrated light suffers from severe biases when extended HB stars are present, if those stars are not included in the modelling. 
In this work we overcome this longstanding roadblock to the determination of accurate ages from integrated light by including the effect of an extended HB population in the stellar population models used to fit high SNR (mean $\mbox{SNR}\sim100, <4100 \mbox{~\AA}$), low resolution (150 km\,s$^{-1}$) optical ({$3750-5500$} \AA) integrated spectra.

In our experiments we compare the age, metallicity and horizontal branch properties of globular clusters derived from the fits to their integrated spectra with the literature values obtained from independent methods, i.e. CMD analysis (for the ages and HB properties) and high resolution spectra (for the metallicities). Although the metallicity recovered from the integrated spectra is in good agreement with the values inferred from high-resolution spectra or individual stars regardless if the extended HB is modelled or not, the inferred age changes significantly.

We find that the ages obtained using stellar population models that include a contribution of extended HB stars yielded ages that are within 0.1 dex of the values obtained from the analysis of their CMD for { $\sim81\%$} the old globular clusters in our sample. On the other hand, when using standard stellar population models (i.e. not including extended HB stars) we find that only minority $\sim28\%$ are within 0.1 dex from their CMD values.  We also present a diagnostic based on the differences between the observed and best fit model spectra that can be used to identify when the ages inferred from the spectral fits are yield spurious age solutions (see Fig. \ref{fig:ca_res}).

From our fits to the integrated spectra we are able to recover the flux contribution of the extended HB population for { $\sim75\%$} of our targets within 0.15 dex of the value inferred from the CMD. Similarly, from our spectral fits we recovered the mean temperature for the extended HB population for { $\sim58\%$} of our targets within { $<30\%$} of the value derived from their CMDs. For the remaining targets, we find that the available integrated spectra might not be a good representation of the stellar population inferred from the targets CMDs. This is possibly due to stochastic sampling in different evolutionary stages due to the small apertures used to obtain the \cite{Schiavon05} spectra, or foreground/background contamination, or a problem with our models.

The performance of the technique demonstrated in this work has opened a new opportunity to carry out the type of analysis that was restricted to nearby resolved stellar populations to distant environments.  In particular, globular clusters are known to be great tracers to study the formation and evolution of galaxies. Their metallicties encode information about the chemical properties of the gas of its host galaxy and their ages tag this information to a specific time in history \citep[e.g.][]{Pfeffer18}. Although this has been used for decades to gain insights about how galaxies grow \citep[e.g.][]{Searle78}, only recently it has become possible to do this in a quantitative way. 

For example, \cite{Kruijssen19} used the (CMD) ages, metallicities and masses of Galactic globular clusters to reconstruct the assembly history of the MW in detail. Although recent analysis of extragalactic globular clusters been used to reveal a diversity in the formation history of different galaxies \citep[e.g.][]{Beasley15,Usher19}, the results are limited by the age uncertainties from integrated light studies.

With 0.1 dex uncertainties in the age of extragalactic globular clusters, we will be able to reconstruct the accretion history of their host galaxies with a precision comparable to the one found recent studies of the Milky Way. Similarly, the analysis of large samples of globular clusters (in order to overcome random errors) with this kind of accuracy could be used to estimate when most of globular clusters where formed, and distinguish  
between scenarios predicting globular cluster formation peaked around $z\sim2$ ($\mbox{log(age/yr)}\sim10.02$) and $z>7$ ($\mbox{log(age/yr)}\gtrsim10.11$) -- see Section \ref{sec:intro}.

Finally, we note there is the potential to improve these results by folding in the information about the SED when modelling the stellar integrated stellar populations like it is done with other algorithms e.g. \cite{Johnson20,Werle20}, as well as implementing a more sophisticated modelling of the extended HB. This will be explored in a future work.

\section*{Acknowledgements}

IC-Z was supported by the European Research Council (ERC-CoG-646928, Multi-Pop) and NASA through the Hubble Fellowship grant HST-HF2-51387.001-A awarded by the Space Telescope Science Institute, which is operated by the Association of Universities for Research in Astronomy, Inc., for NASA, under contract NAS5-26555. This research was also supported in part by the National Science Foundation under Grant No. NSF PHY-1748958. We would like to thank the organisers and participants of the KITP Program: Globular Clusters at the Nexus of Star and Galaxy Formation (Mar. 30 - May. 29, 2020) for the inspiring discussions that led to the development of this work. We also thank C. Usher, R. Schiavon, M. Salaris and N. Bastian for insightful feedback.

This research is based on observations with the NASA/ESA \emph{Hubble Space Telescope} obtained at the Space Telescope Science Institute. The computations in this paper were run on the FASRC Cannon cluster supported by the FAS Division of Science Research Computing Group at Harvard University.

\section*{Data Availability}

The data used in our analysis is publicly available and can be found in the following links:

 \renewcommand{\labelitemi}{\textendash}
 \begin{itemize}
    \item Galactic globular cluster spectra  \citep{Schiavon05}: \url{https://www.noao.edu/ggclib/}
    \item WAGGS spectra \citep{Usher17}: \url{https://www.astro.ljmu.ac.uk/~astcushe/waggs/}
    \item WFPC2 catalogues \citep{Piotto02}: \url{ https://cdsarc.unistra.fr/viz-bin/cat/J/A+A/391/945}
    \item ACS catalogues \citep{Sarajedini07}: \url{https://archive.stsci.edu/prepds/acsggct/}
    \item MIST ischrones \citep{Choi16}: \url{http://waps.cfa.harvard.edu/MIST/}
    \item Zero age HB models \citep{Pietrinferni06}: \url{http://basti.oa-teramo.inaf.it/BASTI/DATA_AE_S/aems_c.php}
    \item Parameters from \cite{Baumgardt19}:
    \url{https://people.smp.uq.edu.au/HolgerBaumgardt/globular/}
\end{itemize}




\bibliographystyle{mnras}
\bibliography{mnras} 




\appendix

\section{Complementary Figures}

\begin{figure*}
	\includegraphics[width=170mm]{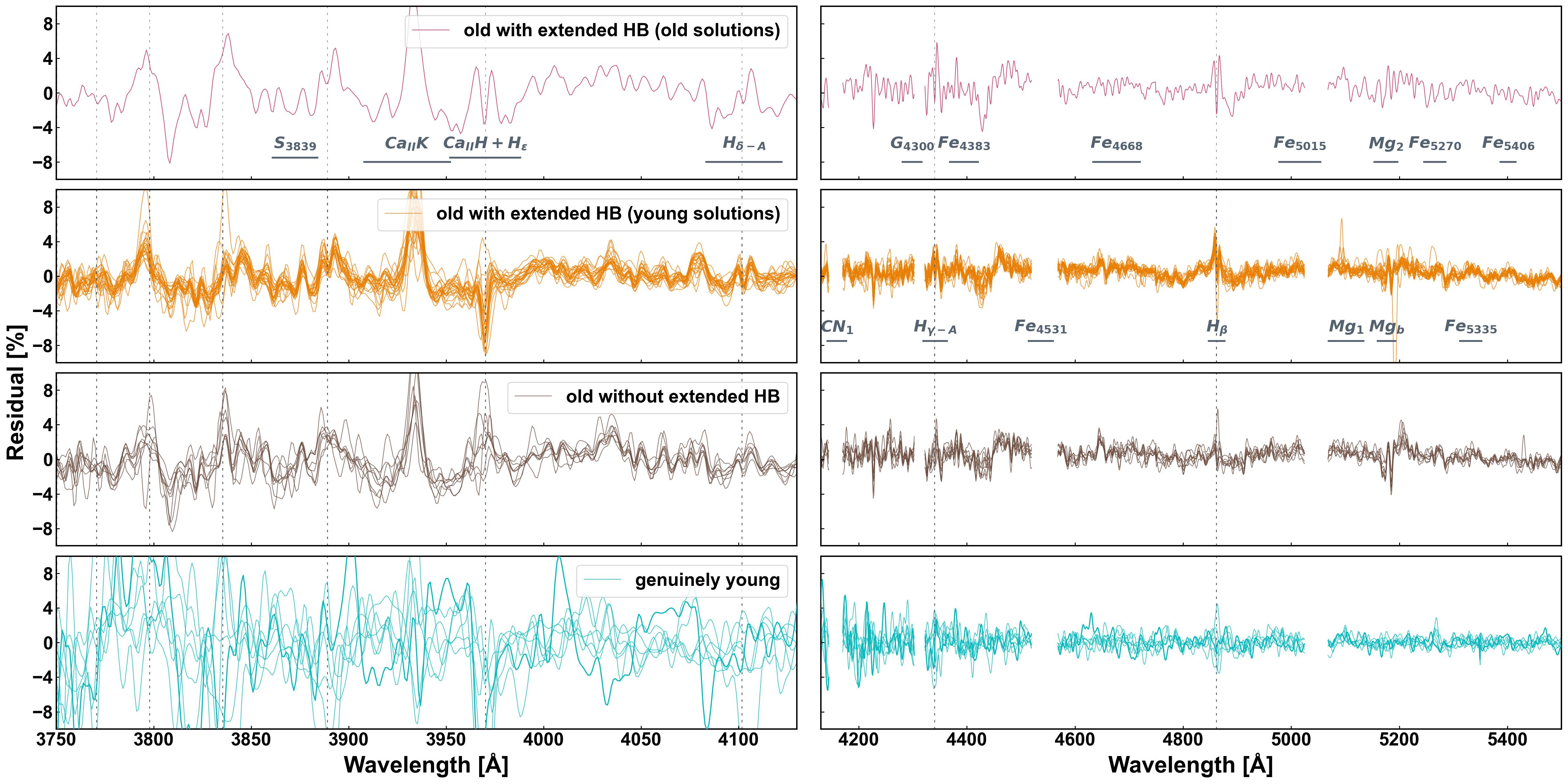}
    \caption{Similar to Fig. \ref{fig:ehb_residuals} but for standard stellar population models.}
    \label{fig:ehb_residuals2}
\end{figure*}

\begin{figure*}
	\includegraphics[width=170mm]{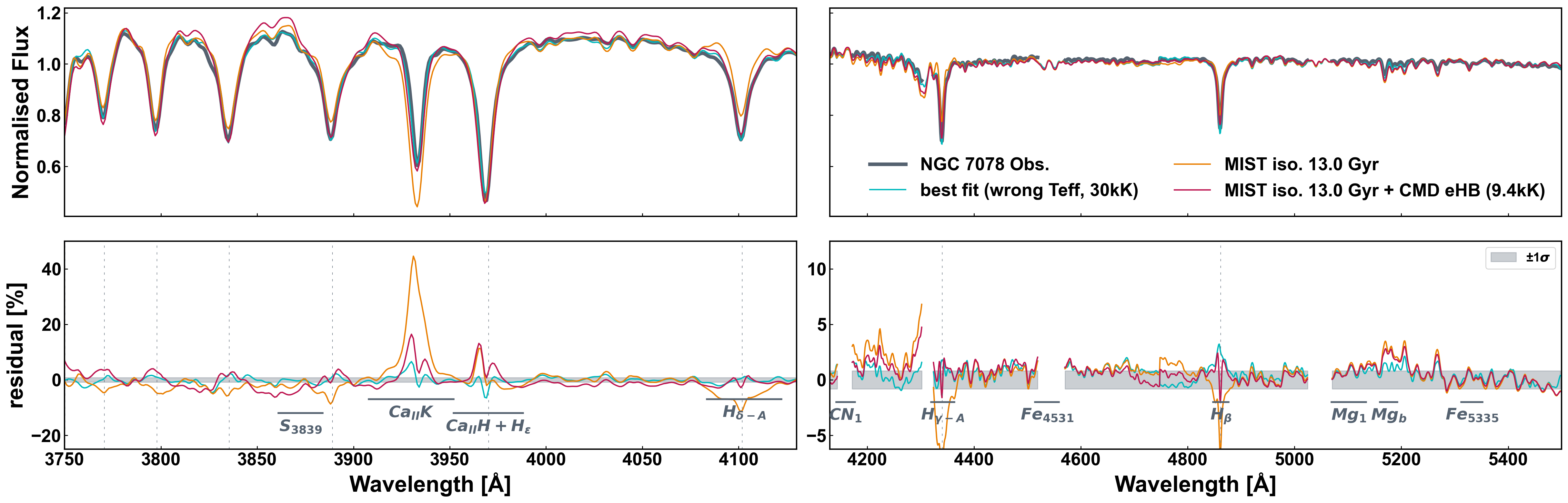}
    \caption{\textbf{Top:} The grey line shows the \citeauthor{Schiavon05} spectrum of NGC 7078. Our best fit model is shown in blue (with a temperature for the extended HB component of $30 {\rm ~kK}$). In yellow we show the integrated spectra given by a MIST isochrone of with the reference age and metallicity. In red we show the integrated spectra given by a MIST isochrone of with the reference age and metallicity plus the contribution of the extended HB population as inferred from the cluster's CMD. \textbf{Bottom:} Residuals (data-model)/model for the different cases. Note that the scale of the residuals is different in the left and right panel.}
    \label{fig:n7078}
\end{figure*}

\begin{figure*}
	\includegraphics[width=170mm]{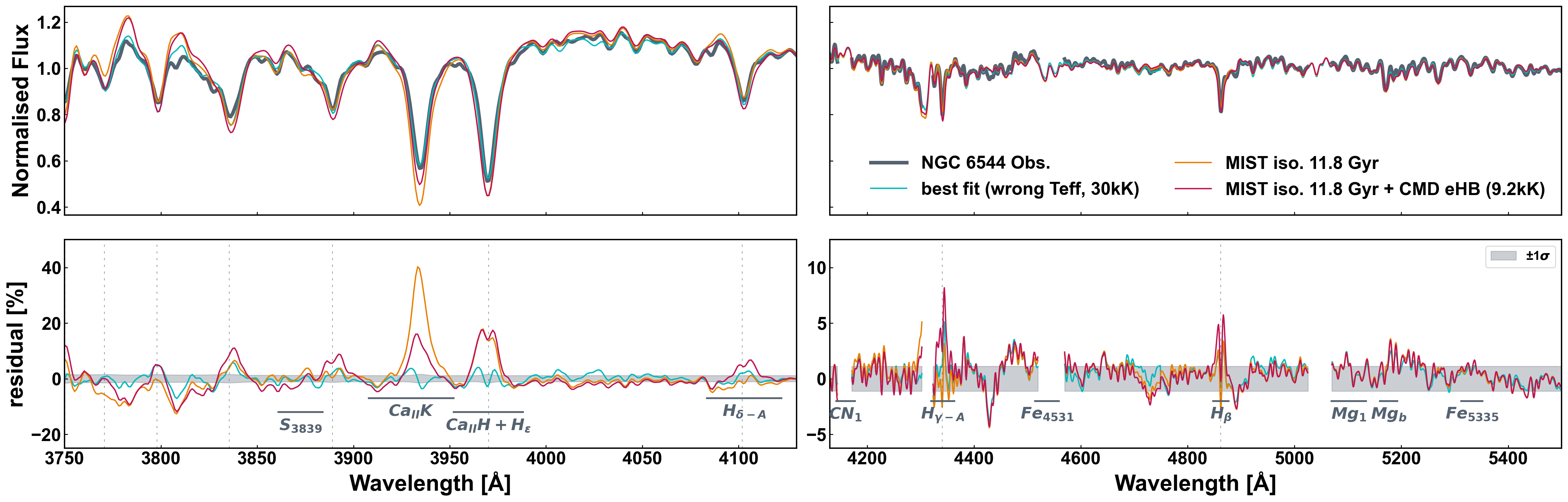}
    \caption{Similar to Fig. \ref{fig:n7078} but for NGC 6544.}
    \label{fig:n6544}
\end{figure*}

\section{Tables}

\begin{table*}
	\centering
	\caption{The first 11 columns contain the literature parameters for the targets in our sample. The following columns contain the parameters inferred in this work using models with extended HB and their $1\sigma$ statistical uncertainties. The units for the age, [Fe/H], $m-{\rm M}_0$, ${\rm A}_V$  and $T_{\rm hot}$ are Gyr, dex, mag, mag and kK respectively. Targets with extended HB are the ones with values in the $m-{\rm M}_0$ and ${\rm A}_V$ columns.}
	\label{tab:lit}
	\resizebox{\textwidth}{!}{
\begin{tabular}{cccccccccccccccccccc}
		\hline
	& & & & & Literature &&&&&&&&&& This work &&&\\	
ID & Age & $\sigma$Age & [Fe/H] & $\sigma$[Fe/H] & $m-{\rm M}_0$ & ${\rm A}_V$ & Age ref. & [Fe/H] ref. & $m-{\rm M}_0$ ref. & ${\rm A}_V$ ref. & Age & $\sigma$Age & [Fe/H] & $\sigma$[Fe/H] & $\log(f_{\rm hot})^\star$ & $\sigma \log(f_{\rm hot})$ & $T_{\rm hot}$ & $\sigma T_{\rm hot}$ & $\chi^2/{\rm d.o.f.}$\\
		\hline
NGC 0104 &       12.5 &  0.6 &   -0.75 &         0.03 &    -   &           -   &         1 &     1 &     - &     -  & 14.00 & 0.03 & -0.79 & 0.01 & -1.63 & 0.02 & 29.99 & 0.06 & 0.697  \\
NGC 1851 &       10.5 &  0.5 &   -1.10 &         0.07 &  15.43 &         0.105 &         1 &     1 &     6 &     6  & 13.96 & 0.39 & -1.25 & 0.01 & -0.84 & 0.01 & 8.01 & 0.04 & 0.262 \\  
NGC 1904 &       11.1 &  0.9 &   -1.37 &         0.20 &  15.61 &         0.031 &         1 &     1 &     2 &     2  & 13.93 & 0.26 & -1.56 & 0.01 & -0.65 & 0.01 & 8.04 & 0.10 & 0.418 \\  
NGC 2298 &       12.8 &  0.2 &   -1.80 &         0.09 &  15.17 &         0.718 &         1 &     1 &     2 &     7  & 5.53 & 0.09 & -1.56 & 0.02 & -0.51 & 0.03 & 27.99 & 0.81 & 0.823 \\  
NGC 2808 &       10.9 &  0.1 &   -1.15 &         0.03 &  15.05 &         0.704 &         1 &     1 &     2 &     6  & 10.05 & 0.21 & -1.13 & 0.01 & -1.02 & 0.01 & 28.98 & 0.59 & 0.323 \\ 
NGC 3201 &       11.2 &  0.7 &   -1.42 &         0.12 &  13.31 &         0.868 &         1 &     1 &     2 &     6  & 12.88 & 0.58 & -1.31 & 0.02 & -0.67 & 0.02 & 8.02 & 0.14 & 1.826 \\  
NGC 5286 &       12.7 &  0.2 &   -1.60 &         0.14 &  15.29 &         0.905 &         1 &     1 &     2 &     6  & 13.03 & 0.24 & -1.55 & 0.01 & -0.65 & 0.01 & 8.02 & 0.07 & 0.395 \\  
NGC 5904 &       11.5 &  0.7 &   -1.25 &         0.09 &  14.40  &        0.118 &         1 &     1 &     2 &     6  & 11.30 & 0.13 & -1.36 & 0.01 & -0.72 & 0.01 & 9.25 & 0.20 & 0.308 \\  
NGC 5927 &       11.9 &  0.8 &   -0.48 &         0.14 &    -   &           -   &         1 &     1 &     - &     -  & 14.00 & 0.06 & -0.49 & 0.01 & -1.38 & 0.02 & 30.00 & 0.05 & 1.070 \\
NGC 5946 &       11.4 &  0.9 &   -1.22 &         0.20 &  15.13 &         1.674 &         1 &     1 &     2 &     2  & 5.93 & 0.10 & -1.40 & 0.02 & -0.69 & 0.03 & 29.90 & 0.53 & 1.090 \\  
NGC 5986 &       12.6 &  0.5 &   -1.53 &         0.13 &  15.12 &         0.894 &         1 &     1 &     2 &     7  & 13.03 & 0.24 & -1.58 & 0.01 & -0.67 & 0.01 & 8.69 & 0.14 & 0.361 \\  
NGC 6121 &       12.2 &  0.5 &   -1.14 &         0.07 &  11.43 &         1.339 &         1 &     1 &     2 &     8  & 12.25 & 0.37 & -1.25 & 0.01 & -0.71 & 0.01 & 8.00 & 0.01 & 0.487 \\  
NGC 6171 &       12.9 &  0.8 &   -0.99 &         0.03 &  13.78 &         1.349 &         1 &     1 &     3 &     6  & 8.67 & 0.24 & -1.01 & 0.02 & -0.87 & 0.02 & 29.84 & 0.10 & 0.627 \\  
NGC 6218 &       13.0 &  0.2 &   -1.26 &         0.08 &  13.29 &         0.651 &         1 &     1 &     2 &     6  & 12.76 & 0.63 & -1.41 & 0.01 & -0.67 & 0.01 & 8.54 & 0.15 & 0.500 \\  
NGC 6235 &       11.4 &  0.9 &   -1.18 &         0.20 &  15.65 &         0.961 &         1 &     1 &     2 &     2  & 13.06 & 0.52 & -1.20 & 0.02 & -0.85 & 0.02 & 10.93 & 0.41 & 0.922 \\
NGC 6254 &       12.0 &  0.7 &   -1.46 &         0.15 &  13.48 &         0.868 &         1 &     1 &     2 &     2  & 11.75 & 0.44 & -1.41 & 0.01 & -0.71 & 0.01 & 14.69 & 0.32 & 0.529 \\
NGC 6266 &       11.8 &  0.9 &   -1.02 &         0.20 &  14.03 &         1.240 &         1 &     1 &     2 &     2  & 11.69 & 0.36 & -1.13 & 0.01 & -0.95 & 0.01 & 14.84 & 0.39 & 0.587 \\
NGC 6284 &       11.1 &  0.9 &   -1.13 &         0.20 &  15.90  &        0.868 &         1 &     1 &     2 &     2  & 11.78 & 0.33 & -1.22 & 0.01 & -0.76 & 0.01 & 11.40 & 0.09 & 0.347 \\
NGC 6304 &       12.5 &  1.0 &   -0.51 &         0.12 &    -   &           -   &         1 &     1 &     - &     -  & 14.00 & 0.03 & -0.55 & 0.01 & -1.47 & 0.02 & 29.99 & 0.06 & 1.185 \\
NGC 6342 &       12.0 &  0.9 &   -0.69 &         0.20 &    -   &           -   &         1 &     1 &     - &     -  & 13.87 & 0.26 & -0.87 & 0.01 & -1.48 & 0.05 & 29.86 & 0.40 & 0.803 \\
NGC 6352 &       12.1 &  1.0 &   -0.71 &         0.07 &    -   &           -   &         1 &     1 &     - &     -  & 14.00 & 0.10 & -0.69 & 0.02 & -1.73 & 0.04 & 29.99 & 0.18 & 1.523 \\
NGC 6362 &       12.9 &  0.5 &   -1.05 &         0.05 &  14.33 &         0.236 &         1 &     1 &     2 &     6  & 9.93 & 0.16 & -1.24 & 0.01 & -1.00 & 0.01 & 29.85 & 0.10 & 0.458 \\  
NGC 6388 &       12.0 &  1.0 &   -0.77 &         0.20 &  15.16 &         1.147 &         1 &     1 &     2 &     2  & 12.27 & 0.26 & -0.68 & 0.01 & -1.26 & 0.01 & 28.16 & 0.73 & 0.648 \\
NGC 6544 &       11.8 &  1.1 &   -1.47 &         0.07 &  11.96 &         2.249 &         18 &    19 &    20 &    20 & 13.68 & 0.22 & -1.25 & 0.02 & -0.64 & 0.01 & 29.98 & 0.03 & 1.259 \\
NGC 6624 &       12.3 &  0.7 &   -0.54 &         0.12 &    -   &           -   &         1 &     1 &     - &     -  & 14.00 & 0.03 & -0.72 & 0.01 & -1.63 & 0.02 & 29.95 & 0.11 & 0.609 \\
NGC 6637 &       12.2 &  0.9 &   -0.69 &         0.08 &    -   &           -   &         1 &     1 &     - &     -  & 14.00 & 0.03 & -0.92 & 0.01 & -1.80 & 0.03 & 29.99 & 0.09 & 0.708 \\
NGC 6638 &       12.0 &  0.5 &   -0.99 &         0.07 &  15.07 &         1.271 &         5 &    19 &     2 &     2  & 12.11 & 0.37 & -1.02 & 0.01 & -1.06 & 0.01 & 29.97 & 0.17 & 0.621 \\
NGC 6652 &       12.5 &  0.9 &   -0.83 &         0.10 &    -   &           -   &         1 &     1 &     - &     -  & 13.90 & 0.19 & -0.92 & 0.01 & -1.42 & 0.02 & 8.01 & 0.05 & 0.511 \\  
NGC 6723 &       12.8 &  0.2 &   -1.02 &         0.06 &  14.67 &         0.221 &         1 &     1 &     4 &     4  & 11.12 & 0.29 & -1.25 & 0.01 & -1.01 & 0.03 & 8.79 & 0.28 & 0.506 \\  
NGC 6752 &       12.3 &  0.3 &   -1.43 &         0.14 &  13.14 &         0.124 &         1 &     1 &     2 &     2  & 13.06 & 0.27 & -1.55 & 0.02 & -0.61 & 0.01 & 8.40 & 0.15 & 0.475 \\  
NGC 7078 &       13.0 &  0.2 &   -2.25 &         0.17 &  15.05 &         0.335 &         1 &     1 &     2 &     6  & 5.71 & 0.05 & -1.86 & 0.01 & -0.31 & 0.02 & 29.96 & 0.09 & 0.662 \\  
NGC 7089 &       12.0 &  0.3 &   -1.52 &         0.15 &  15.11 &         0.143 &         1 &     1 &     2 &     6  & 12.85 & 0.54 & -1.52 & 0.01 & -0.61 & 0.01 & 8.21 & 0.14 & 0.372 \\  
Kron 3   &       6.5  &  0.5 &   -1.08 &         0.12 &    -   &           -   &        10 &    11 &     - &     - & 8.95 & 0.57 & -1.22 & 0.02 & -1.72 & 0.35 & 8.09 & 3.17 & 0.051 \\  
NGC 0411 &       1.4  &  0.1 &   -0.70 &         0.10 &    -   &           -   &         9 &     9 &     - &     - & 2.21 & 0.04 & -0.69 & 0.01 & -3.89 & 0.42 & 27.83 & 6.55 & 0.048 \\ 
NGC 0416 &       6.0  &  0.5 &   -1.00 &         0.13 &    -   &           -   &        10 &    12 &     - &     - & 6.68 & 0.17 & -1.17 & 0.01 & -1.56 & 0.06 & 8.00 & 0.02 & 0.020 \\  
NGC 0419 &       1.5  &  0.1 &   -0.77 &         0.10 &    -   &           -   &         9 &     9 &     - &     - & 2.32 & 0.19 & -0.63 & 0.01 & -0.50 & 0.09 & 8.04 & 0.04 & 0.056 \\  
NGC 1831 &       0.8  &   -  &   -0.49 &         -    &    -   &           -   &        13 &    13 &     - &     - & 0.81 & 0.03 & -0.52 & 0.02 & -0.01 & 0.02 & 9.64 & 0.14 & 0.095 \\  
NGC 1868 &       1.1  &  0.1 &   -0.50 &          -   &    -   &           -   &        14 &    15 &     - &     - & 1.70 & 0.05 & -0.51 & 0.01 & -0.15 & 0.01 & 8.00 & 0.00 & 0.026 \\  
NGC 1978 &       1.9  &  0.1 &   -0.38 &         0.07 &    -   &           -   &        16 &    17 &     - &     - & 3.01 & 0.02 & -0.55 & 0.01 & -1.77 & 0.04 & 25.39 & 1.03 & 0.028 \\ 
		\hline
\multicolumn{19}{l}{$^\star$ (log) fraction of flux of extended HB stars to the integrated light at 0.5\micron.}\\
\multicolumn{19}{l}{1) Average from Galactic studies see Section \ref{sec:phot_ages}; 2) distance from \cite{Baumgardt19} reddenning from \cite{Harris96} (2010 edition); 3) \cite{OConnell11}; 4) \cite{Lee14}; 5) \cite{Meissner06};}\\
\multicolumn{19}{l}{6) \cite{VandenBerg13}; 7) \cite{Dotter10}; 8) $R_V\neq3.1$ \cite{Hendricks12}; 9) \cite{Goudfrooij14}; 10) \cite{Glatt08}; 11) \cite{DaCosta98}; 12) \cite{Glatt09};}  \\
\multicolumn{19}{l}{13) \cite{Correnti21}; 14) \cite{Kerber07}; 15) \cite{Olszewski91}; 16) \cite{Mucciarelli07}; 17) \cite{Ferraro06}; 18) age average between \cite{Forbes10} and \cite{Cohen14}}\\
\multicolumn{19}{l}{19) \cite{Carretta09}; 20) \cite{Cohen14}}\\
	\end{tabular}}
\end{table*}

\begin{table}
\centering
\caption{Results of spectral fits using standard models. The units for age and [Fe/H] are Gyr and dex respectively.}
\label{tab2}
\begin{tabular}{cccccc}
\hline
ID & Age & $\sigma$Age & [Fe/H] & $\sigma$[Fe/H] & $\chi^2/{\rm d.o.f.}$\\
\hline
NGC 0104 &  13.96 & 0.19 & -0.82 & 0.02 & 1.013 \\
NGC 1851 &  6.01  & 0.13 & -1.07 & 0.03 & 0.585 \\
NGC 1904 &  5.65  & 0.12 & -1.35 & 0.03 & 0.890 \\
NGC 2298 &  5.51  & 0.13 & -1.57 & 0.03 & 1.025 \\
NGC 2808 &  7.98  & 0.24 & -1.10 & 0.02 & 0.640 \\
NGC 3201 &  5.31  & 0.15 & -1.14 & 0.03 & 2.152 \\
NGC 5286 &  5.75  & 0.12 & -1.41 & 0.03 & 0.816 \\
NGC 5904 &  4.53  & 0.08 & -1.08 & 0.03 & 0.842 \\
NGC 5927 &  11.27 & 0.45 & -0.52 & 0.02 & 1.821 \\
NGC 5946 &  5.94  & 0.14 & -1.38 & 0.03 & 1.331 \\
NGC 5986 &  5.52  & 0.13 & -1.38 & 0.03 & 0.792 \\
NGC 6121 &  4.82  & 0.20 & -1.05 & 0.03 & 0.829 \\
NGC 6171 &  7.73  & 0.31 & -1.04 & 0.03 & 0.971 \\
NGC 6218 &  4.55  & 0.08 & -1.14 & 0.02 & 1.068 \\
NGC 6235 &  5.61  & 0.21 & -1.08 & 0.04 & 1.171 \\
NGC 6254 &  6.25  & 0.13 & -1.34 & 0.03 & 0.997 \\
NGC 6266 &  6.21  & 0.14 & -0.99 & 0.03 & 1.102 \\
NGC 6284 &  4.60  & 0.10 & -0.98 & 0.03 & 1.069 \\
NGC 6304 &  13.71 & 0.42 & -0.63 & 0.02 & 2.000 \\
NGC 6342 &  11.51 & 0.65 & -0.90 & 0.03 & 0.979 \\
NGC 6352 &  13.80 & 0.42 & -0.74 & 0.02 & 1.769 \\
NGC 6362 &  8.71  & 0.33 & -1.18 & 0.03 & 0.854 \\
NGC 6388 &  6.71  & 0.24 & -0.52 & 0.02 & 1.102 \\
NGC 6544 &  11.09 & 0.39 & -1.36 & 0.03 & 2.078 \\
NGC 6624 &  13.87 & 0.42 & -0.78 & 0.02 & 0.976 \\
NGC 6637 &  13.96 & 0.29 & -0.92 & 0.02 & 0.961 \\
NGC 6638 &  8.81  & 0.35 & -0.98 & 0.03 & 1.367 \\
NGC 6652 &  10.14 & 0.31 & -0.88 & 0.02 & 0.752 \\
NGC 6723 &  6.64  & 0.19 & -1.09 & 0.03 & 0.887 \\
NGC 6752 &  5.22  & 0.18 & -1.34 & 0.04 & 1.042 \\
NGC 7078 &  5.92  & 0.10 & -1.79 & 0.04 & 1.041 \\
NGC 7089 &  5.37  & 0.18 & -1.31 & 0.03 & 0.900 \\
Kron 3   & 6.82 & 2.11 & -1.22 & 0.20 & 0.055 \\
NGC 0411 & 1.75 & 0.41 & -0.69 & 0.20 & 0.054 \\
NGC 0416 & 5.62 & 0.88 & -1.18 & 0.13 & 0.028 \\
NGC 0419 & 1.49 & 0.15 & -0.64 & 0.12 & 0.062 \\
NGC 1831 & 0.56 & 0.08 & -0.60 & 0.17 & 0.127 \\
NGC 1868 & 1.10 & 0.22 & -0.50 & 0.21 & 0.036 \\
NGC 1978 & 3.57 & 0.81 & -0.65 & 0.14 & 0.033 \\
\hline
\end{tabular}
\end{table}



\bsp	
\label{lastpage}
\end{document}